%% file: main.tex
\pdfoutput=1

\documentclass[12pt]{book}
\usepackage[utf8]{inputenc} 
\usepackage{a4wide}
\usepackage{fancyhdr}
\usepackage{graphicx}
\usepackage[bookmarks=false]{hyperref}

\usepackage{amsfonts, amsmath, amssymb}
\usepackage[makeroom]{cancel}
\usepackage{mathtools}
\usepackage[none]{hyphenat}
\usepackage{subfigure}

\usepackage{appendix}
\usepackage{float}
\usepackage[nottoc, notlot, notlof]{tocbibind}
\usepackage[round]{natbib}

\renewcommand{\chaptermark}[1]%
         {\markboth{\thechapter.\ #1}{}}
\renewcommand{\sectionmark}[1]%
         {\markright{\thesection\ #1}}
\newcommand{\LMUTitle}[9]{
  \thispagestyle{empty}

  \begin{center}

    \bfseries\LARGE #1\\ 

  \end{center}
  \vspace{1em}
  \begin{center}
    \includegraphics[width=3in]{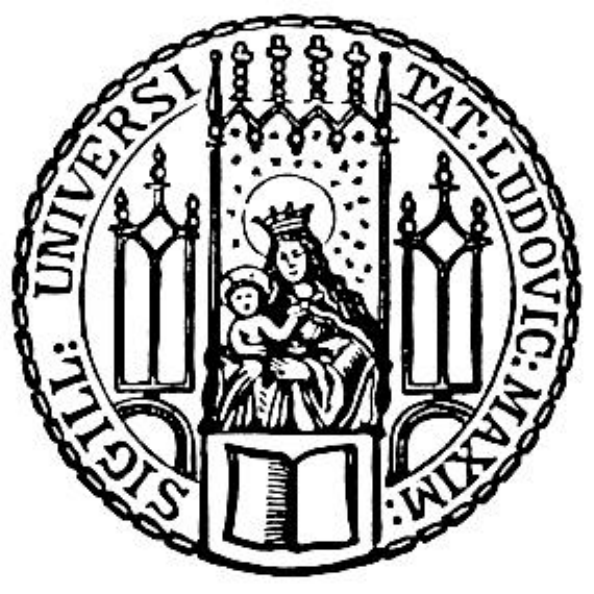}
  \end{center}
  
  \begin{center}
    \large Master Thesis at the #3\\
    \vspace{-0.2cm}
    \large Ludwig-Maximilians-Universität Munich\\
    \vspace*{\stretch{1}}
    \large Submitted by\\
    \vspace{-0.2cm}
    {\bfseries\Large #2}\\
    \vspace*{\stretch{1}}
    \large Supervised by\\
    \vspace{-0.2cm}
    \large #5\\
    \vspace*{\stretch{2}}
    \large Munich, #4
  \end{center}

  \newpage
  \thispagestyle{empty}

  \cleardoublepage
  \thispagestyle{empty}

    \begin{center}

    \bfseries\LARGE #6\\ 

  \end{center}
  \vspace{1em}
  \begin{center}
    \includegraphics[width=3in]{Pictures/siegel.pdf}
  \end{center}
  
  \begin{center}
    \large Masterarbeit an der #7\\
    \vspace{-0.2cm}
    \large Ludwig-Maximilians-Universität München\\
    \vspace*{\stretch{1}}
    \large Eingereicht von\\
    \vspace{-0.2cm}
    {\bfseries\Large #2}\\
    \vspace*{\stretch{1}}
    \large Betreut von\\
    \vspace{-0.2cm}
    \large #5\\
    \vspace*{\stretch{2}}
    \large München, #8
  \end{center}

  \newpage
  \thispagestyle{empty}

  \cleardoublepage
  \thispagestyle{empty}
  
}

\begin{document}

\frontmatter

  \LMUTitle
      {Stacking the Spectra of eROSITA Galaxy Cluster Data for Searches of the 3.5keV line:\\Dark Matter Decay or Charge Exchange?}               
      {Justo Antonio Gonzalez Villalba}                       
      {Faculty of Physics}                         
      {3th of January 2022}                 
      {Dr. Esra Bulbul\\ 
      \vspace{-0.2cm}
      PD Dr. Klaus Dolag}    
      {Kombinierte Spektren von Galaxien Haufen aus eROSITA-daten zur Untersuchung der 3.5keV-Linie: Zerfall der Dunklen Materie oder Ladungsaustausch?}               
      {Fakultät für Physik}                         
      {den 3. Januar 2022}                 

  \tableofcontents

\mainmatter\setcounter{page}{1}

\include{Introduction}

\include{DataProcessing}

\include{Calibration}

\include{Stack}

\include{SpectralAnalysis}

\include{Conclusions}

\backmatter
\include{Bibliography}

\listoffigures

\listoftables

\cleardoublepage

\mainmatter

\backmatter

\include{Declaration}

\end{document}

%% file: Introduction.tex
\chapter{Introduction} \label{sec:Introduction}

\section{Galaxy clusters and the Intra Cluster Medium}

Galaxy clusters are the most massive structures of the Universe, with typical masses ranging $10^{14}-10^{15} M_\odot$. The mass budget can be broken down as follows: 

\begin{itemize}
  \item ~1\% corresponds to galaxies (typically 100 to 1,000 galaxies)
  \item ~9\% corresponds to gas in plasma state, present in between galaxies and also known as Intra Cluster Medium (ICM)
  \item 90\% corresponds to dark matter, whose gravitational influence can be inferred from the velocity dispersion of galaxies, as first discovered by \cite{zwicky1951coma} and also inferred from the temperatures of the plasma with the assumption of Hydrostatic equilibrium. For a comprehensive study using Chandra data see \cite{vikhlinin2006chandra}
\end{itemize}

The Intra Cluster Medium (ICM) is enriched via processes such as Type Ia and core-collapse supernovae that explode, and transport elements over the entire cluster. For a recent review of the ICM enrichment process from the observational and simulations perspectives see \cite{mernier2018enrichment} and \cite{biffi2018enrichment}. This enrichment is measurable via the X-Ray line emission of the various ions, and with enough exposure is possible to measure the abundances of each individual element, including faint emission from weak lines.

\section{Modelling the X-Ray emission of the ICM} \label{sec:APEC}

\begin{figure}[ht!]
     \centering
     \begin{subfigure}
         \centering
         \includegraphics[width=0.45\textwidth]{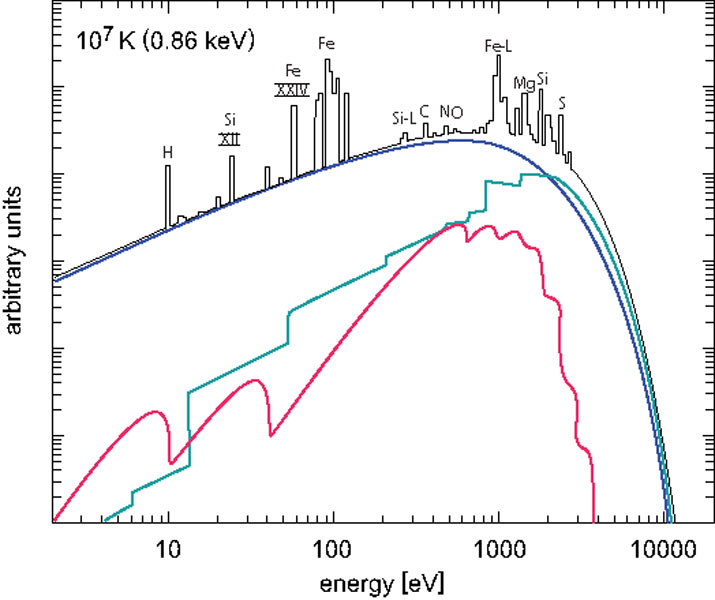}
     \end{subfigure}
     \hfill
     \begin{subfigure}
         \centering
         \includegraphics[width=0.45\textwidth]{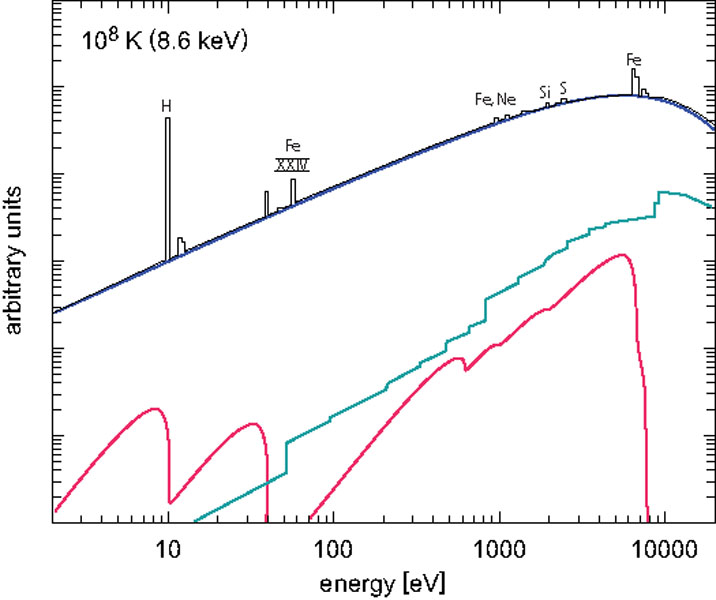}
     \end{subfigure}
     \caption{ \footnotesize { Contributions to the X-ray spectra at two different plasma temperatures: 0.88 keV (left) and 8.6 keV (right). The blue line shows the contribution from bremsstrahlung, the green line shows the contribution from recombination radiation, when an electron is re-capture by an ion, and the red line shows the contribution from de-excitation radiation, when an electron goes from a high energy to a lower energy quantum level. Original from \cite{bohringer2010x} } }
     \label{fig:xray_emission}
\end{figure}

As described in the \cite{bohringer2010x} seminar paper and as shown in Figure \ref{fig:xray_emission}, the X-Ray emission of the intra galaxy-cluster medium (ICM) can be accurately modelled with the following contributions from radiative processes:

\begin{itemize}

  \item Bremsstrahlung: A free-free process in which the trajectory of a free electron is deflected during a fly-by close to an ion. The spectral energy distribution for the thermal bremsstrahlung spectrum for the collision of an electron with ion, i, is given by Equation \ref{eqn:Bremsstrahlung}. Where me and $n_e$ are the electron mass and density, respectively, $n_i$ is the respective ion density, Z is the effective charge of the ion, and $g_{ff}$ is the gaunt factor, a quantity close to unity which must be calculated numerically through a quantum-mechanical treatment.

\begin{equation}
\epsilon(v)=\frac{16 e^{6}}{3 m_{e} c^{2}}\left(\frac{2 \pi}{3 m_{e} k_{B} T_{X}}\right)^{1 / 2} n_{e} n_{i} Z^{2} g_{f f}\left(Z, T_{X}, v\right) \exp \left(\frac{-h v}{k_{B} T_{X}}\right)
\label{eqn:Bremsstrahlung}
\end{equation}
  
  \item Radiation: A free-bound process in which a free electron is re-captured by an ion.
  
  \item De-excitation: A bound-bound process in which an electron changes from a high energy quantum level to a lower energy quantum level
  
\end{itemize}

Such X-Ray emission models are implemented in packages such as the Astrophysical Plasma Emission Code (APEC) \citep{smith2001collisional}, built on top of the Astrophysical Plasma Emission Database (APED),  which provides atomic data such as ionization ratios, collisional and radiative rates and recombination cross sections. The APEC/APED model has been widely accepted by the X-Ray community as a standard to model the X-Ray emission of the ICM.

\section{Detection of an unidentified line signal at ~3.5keV} \label{sec:3.5keV}

\begin{figure}[ht!]
    \centering
    \includegraphics[width=1.0\textwidth]{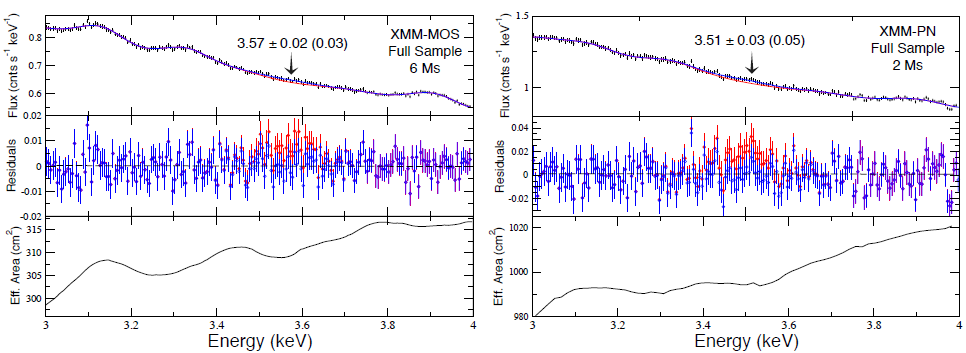}
    \caption{ \footnotesize { Stacked spectra corresponding to MOS CCD camera (left) and PN CCD camera (right).The red bars in the residual subplot show the excess at around 3.5keV if only a 4-temperature apec model is used, whereas the blue bars in the residual subplot show how the excess disappears when adding a Gaussian line at $E = 3.57 \pm 0.02(0.03)$ in the stacked MOS spectra (left) and at $E = 3.51 \pm 0.03(0.05)$ in the stacked PN spectra (right). Original from \cite{bulbul2014detection} }
     \label{fig:bulbul2014detection} }
\end{figure}

\cite{bulbul2014detection} used a technique to stack the spectra of 73 clusters obtained with the MOS and PN CCD cameras aboard XMM-Newton, to reach higher signal-to-noise ratio and potentially detect faint emission features. After modelling the stacked spectra with a 4-temperature APEC model an unidentified line feature (ULF) become apparent at $E = (3.55-3.57) \pm 0.03 keV$ with a $>3 \sigma$ statistical significance - see Figure \ref{fig:bulbul2014detection}.

Also an unidentified line feature at $E = 3.518_{-0.022}^{+0.019} keV$ was detected independently by \cite{boyarsky2014unidentified} in the combined spectra of the Andromeda Galaxy and the Perseus Galaxy Cluster obtained with XMM-Newton, with a $4.4 \sigma$ statistical significance.

\begin{figure}[ht!]
    \centering
    \includegraphics[width=0.6\textwidth]{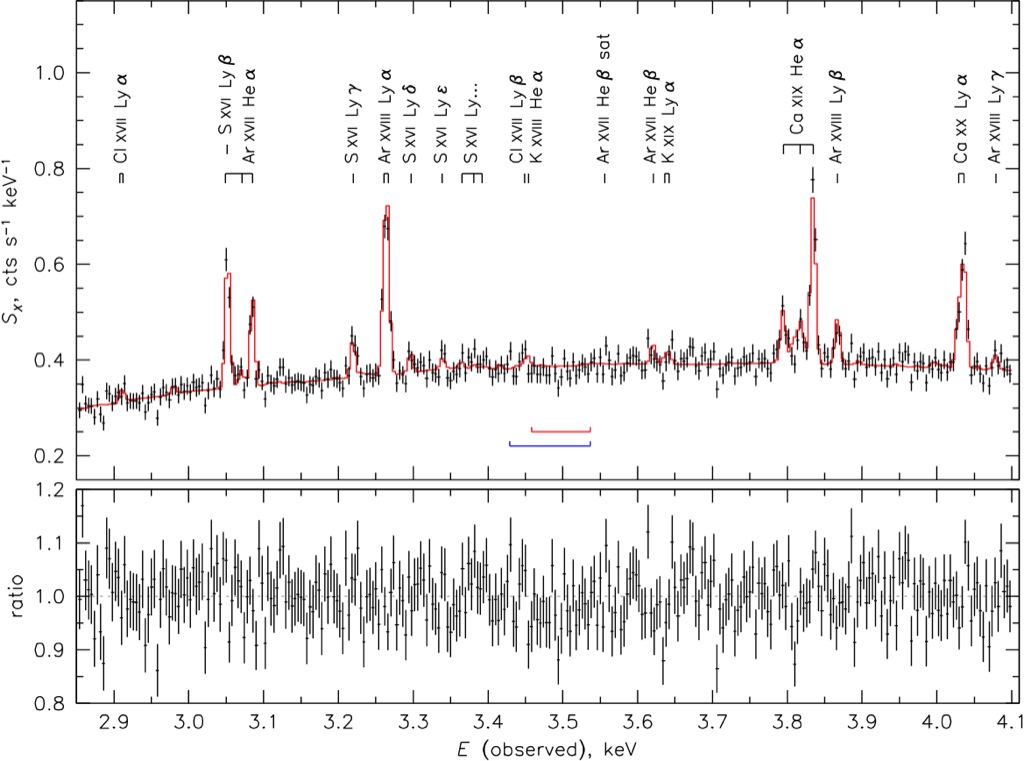}
    \caption{ \footnotesize {X-Ray spectra from Perseus cluster with a total exposure of 275 ks. Spectral resolution is 2 ev in the original data, but the plot/fit is shown with 4 ev energy bins. Error bars show 1-sigma uncertainties using Poisson distribution. The fit (red) curve shows a velocity-broaden apec model (bapec) with temperature 3.5 keV, and a velocity dispersion of 180 km/s. Original from \cite{aharonian2017hitomi} } }
    \label{fig:aharonian2017hitomi}
\end{figure}

On the other hand, non detections have been reported, most notably in the high resolution spectra of the Perseus cluster observed with Hitomi \citep{aharonian2017hitomi} - see Figure \ref{fig:aharonian2017hitomi}, and in the stacked spectra of 89 Galaxies using Chandra and XMM-Newton MOS data \citep{anderson2015non}.

\section{Constrains on the mass of dark matter particles} \label{sec:Constrains}

Since \cite{tremaine1979dynamical} introduces the Tremaine-Gunn bound, there has been an effort to constrain the mass of the dark patter particles. The argument of  \cite{tremaine1979dynamical} was along the lines that if the dark matter particles are Fermions, then the maximum phase space density cannot exceed that of a Fermi-Dirac distribution given by Equation \ref{eqn:Fermi-Dirac}. On the other hand, and isothermal dark matter halo has a maximum phase space density given by the Maxwell-Boltzmann distribution \ref{eqn:Maxwell-Bolzmann}, which depends on the radius. 

\begin{equation}
\begin{aligned}
f(\boldsymbol{p})=\frac{2 h_{P}^{-3}}{\exp \left(p c / k_{B} T_{0}\right)+1} \rightarrow f_{max}(\boldsymbol{p=0}) = h_{P}^{-3}
\end{aligned}
\label{eqn:Fermi-Dirac}
\end{equation}

\begin{equation}
\begin{aligned}
f(\boldsymbol{r}, \boldsymbol{p})=\left(2 \pi m_{\nu}^{2} \sigma^{2}\right)^{-3 / 2} n(r) \exp \left(\frac{-p^{2}}{2 m_{\nu}^{2} \sigma^{2}}\right) \rightarrow f_{max}(\boldsymbol{r},\boldsymbol{p=0}) = \left(2 \pi m_{\nu}^{2} \sigma^{2}\right)^{-3 / 2} n(r)
\end{aligned}
\label{eqn:Maxwell-Bolzmann}
\end{equation}

Equating the maximum phase space densities of the Fermi-Dirac distribution and the Maxwell-Boltzmann distribution, and taking into account that the density profile of an isothermal halo is given by $\rho(r)=\frac{\sigma^{2}}{2 \pi G r^{2}}$, we arrive at Equation \ref{eqn:Tremaine-Gunn} giving a lower limit for the dark matter particle, as a function of radius for a given velocity dispersion $\sigma$. 

\begin{equation}
\begin{aligned}
m_{\nu}>(2 \pi)^{-5 / 8}\left(G h_{P}^{3} \sigma r^{2}\right)^{-1 / 4}
\end{aligned}
\label{eqn:Tremaine-Gunn}
\end{equation}

Applying the Tremaine-Gunn limit to dwarf galaxies yields a lower limit for the mass of the dark matter particles. According to a review by \cite{boyarsky2009lower} this ranges from 715eV (Sextants) to 3.16keV (Canes Venatici II), however more conservative estimates from Leo IV yield 2.19keV, or 1.79keV taking into account the primordial velocity distribution of dark matter particles. 

\section{Dark matter decay scenario: Sterile neutrinos} \label{sec:SterileNeutrino}

Both \cite{bulbul2014detection} and \cite{boyarsky2014unidentified} proposed that such an unidentified line feature could be compatible with a dark matter decay model, namely sterile neutrino decay, which is a natural extension of the standard model, for a recent review see \cite{abazajian2017sterile}.

In the standard model only left-handed neutrinos are considered, as only left-handed neutrinos couple via the electroweak force. If right-handed  neutrinos exist they would not couple to the electroweak force, and thus would be hardly detectable. However, according to the standard model, pure neutrino states do not propagate freely, therefore right-handed neutrinos ($\nu_{RH}$) could mix with left-handed neutrinos ($\nu_{LH}$) resulting in effective mixed neutrino states ($\nu_{\alpha}$) and ($\nu_{\beta}$) as shown in equation \ref{eqn:sterile_neutrino_mixing}:

\begin{equation}
\begin{aligned}
&\left|\nu_{\alpha}\right\rangle=\cos \theta\left|\nu_{LH}\right\rangle+\sin \theta\left|\nu_{RH}\right\rangle \\
&\left|\nu_{\beta}\right\rangle=-\sin \theta\left|\nu_{LH}\right\rangle+\cos \theta\left|\nu_{RH}\right\rangle
\end{aligned}
\label{eqn:sterile_neutrino_mixing}
\end{equation}

Such mixing would thus enable a channel for right-handed neutrinos to interact, and possibly decay in 2 photons whose summed energy accounts for the mass of the sterile neutrino. As shown in Equation \ref{eqn:sterile_neutrino_decay} the decay rate can be parametrized with the mass of the sterile neutrino $m_{s}$ and with the mixing angle $2 \theta$ that accounts for the mixing between right-handed and left-handed neutrinos,  between right-handed and left-handed neutrinos:

\begin{equation}
\Gamma_{\gamma}\left(m_{s}, \theta\right)=1.38 \times 10^{-29} \mathrm{~s}^{-1}\left(\frac{\sin ^{2} 2 \theta}{10^{-7}}\right)\left(\frac{m_{s}}{1 \mathrm{keV}}\right)^{5}
    \label{eqn:sterile_neutrino_decay}
\end{equation}

\begin{figure}[ht!]
    \centering
    \includegraphics[width=1.0\textwidth]{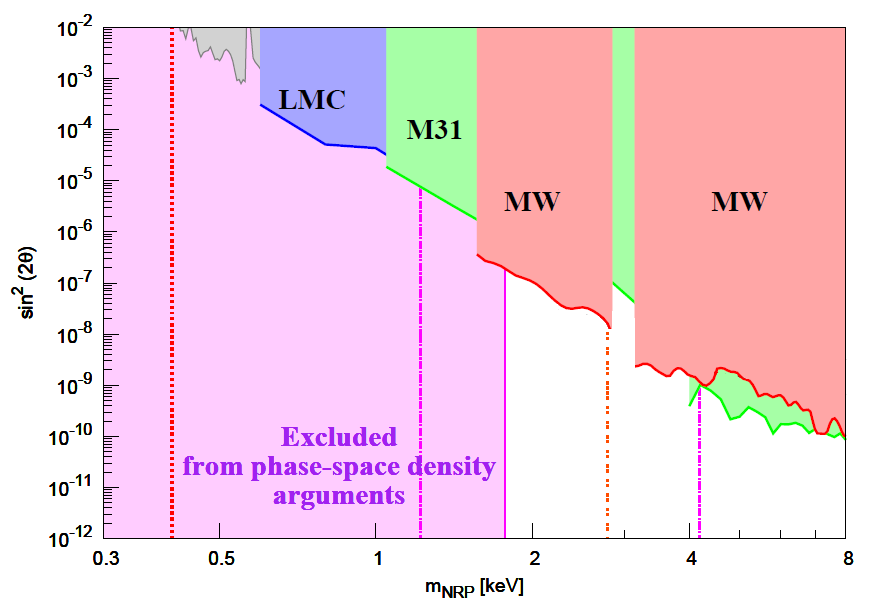}
    \caption{ \footnotesize { Restrictions on parameters of sterile neutrino (mass and mixing $\sin ^{2}(2 \theta)$ between sterile and active neutrinos) and phase-space density considerations. Original from \cite{boyarsky2009lower} } }
    \label{fig:DM-Constrains}
\end{figure}

Additionally, the combination of constrains from phase space density described in \ref{sec:Constrains}, and X-Ray emission from the nearest objects (Milky Way, M31, Large Magellanic Clouds) assuming a sterile neutrino decay, leads to an overall constrained space as shown in Figure \ref{fig:DM-Constrains}, where the 3.5keV band lays in the middle of the allowed parameter space.

Notice that the decay rates of the sterile neutrino scenario are still compatible with the large scale structures observed in the Universe. For instance, assuming the values obtained from the whole sample by \cite{bulbul2014detection}, which are $m_{s}=2 E=7.1 \mathrm{keV}$ for the sterile neutrino mass, and $\sin ^{2}(2 \theta) \approx 7 \times 10^{-11}$ for the mixing angle, one obtains using via \ref{eqn:sterile_neutrino_decay} a decay rate of $1.74 \cdot 10^{-28} s^{-1}$, which corresponds to a half life time of about 10 orders of magnitude higher than the Hubble time as shown in Equation \ref{eqn:hubble_time}:

\begin{equation}
    \begin{aligned}
    &t_{1 / 2}=\tau \ln (2)= 3.98 \cdot 10^{27} \mathrm{s} \\
    &t_{H} \equiv \frac{1}{H_{0}}=\frac{1}{67.8(km / s) / Mpc}=4.55 \cdot 10^{17} s
    \end{aligned}
    \label{eqn:hubble_time}
\end{equation}

\begin{figure}[ht!]
    \centering
    \includegraphics[width=0.6\textwidth]{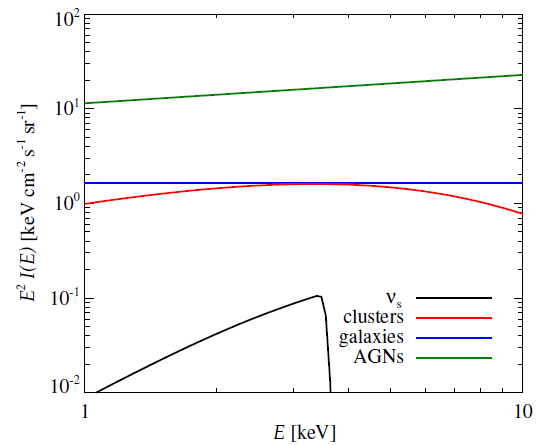}
    \caption{ \footnotesize {Contributions to the cosmic X-Ray background from sterile neutrino decay compared with other astrophysical contributions: unresolved AGNs and galaxies, and galaxy clusters (both resolved and unresolved). Original from \cite{Zandanel_2015} } }
     \label{fig:Zandanel_2015}
\end{figure}

Finally, it is also important to notice that the effect of a sterile neutrino decay in the X-Ray background would be 2 orders of magnitude smaller than the main background contribution from unresolved AGN, as reported by \cite{Zandanel_2015} and shown in Figure \ref{fig:Zandanel_2015}.

\section{Charge exchange scenario: Bare sulfur ions} \label{sec:ChargeExchange}

An alternative explanation to the unidentified line feature at around ~3.5keV was later proposed by \cite{gu2015novel} who suggested that a charge-exchange (CX) process involving bare sulfur ions (SXVI) could also explain an excess emission at around 3.5keV. Charge exchange emission is known to exist, and it was first detected in the the X-Ray emission of comets in the Solar System - for a review see \cite{dennerl2010charge}. However charge exchange emission is not included in the plasma codes used to model the ICM such as APEC/APED.

\begin{figure}[ht!]
    \centering
    \includegraphics[width=0.8\textwidth]{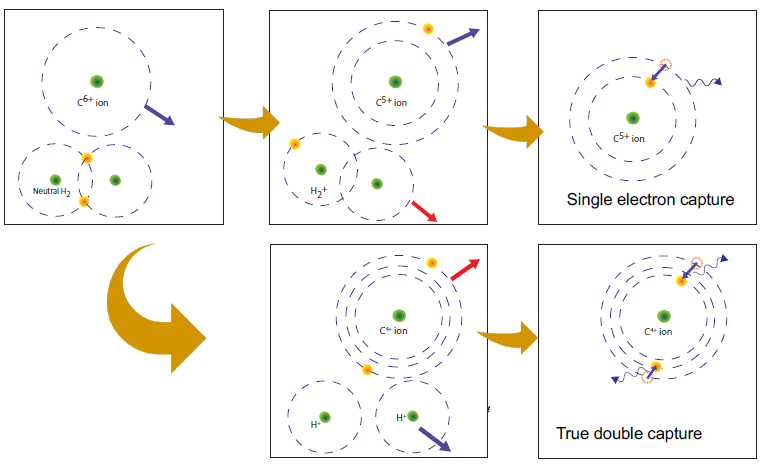}
    \caption{\footnotesize { Example of single (top panels) and double (lower panels) charge exchange reaction: A $C^{6+}$ ion captures one or two electrons from neutral molecular hydrogen. The electrons are placed in high energy states which then cascade down to lower energy states producing X-Ray emission. Original from \cite{cumbee2021interactive} } }
     \label{fig:cumbee2021interactive}
\end{figure}

As shown in Equation \ref{eqn:charge_exchange_reaction} and also depicted in Figure \ref{fig:cumbee2021interactive} the charge exchange process consist of the transfer of an electron from neutral material such as neutral Hydrogen or Helium to an ion. In the context of the ICM a charge exchange scenario would require cold neutral gas (e.g. cold dense clouds) bombarded by ionized elements. The electron then cascades from high energy levels to lower energy levels, producing X-Ray emission lines including those typically forbidden in bound-bound emission due to angular momentum conservation. Such lines are possible in a charge exchange process because the external contribution of angular momentum in the collision.

\begin{equation}
    X^{q+}+Y \rightarrow X^{q-1}\left(n l^{2 S+1} L\right)+Y^{+}
    \label{eqn:charge_exchange_reaction}
\end{equation}

\section{Goals and strategy with eROSITA data} \label{sec:GoalsStrategy}

\begin{figure}[ht!]
    \centering
    \includegraphics[width=0.60\textwidth]{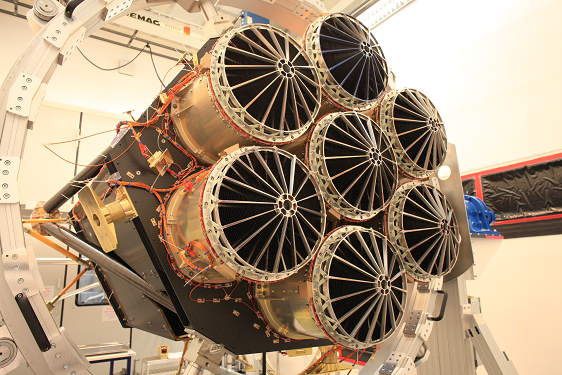}
    \caption { \footnotesize { Front view of eROSITA with all seven mirror assemblies installed. Original from \cite{eROSITA2021} } }
     \label{fig:eROSITA2021_TMs}
\end{figure}

The goal of this work are first to see if an excess at around 3.5keV is detected in the stacked spectra of eROSITA galaxy clusters, and second to determine the nature of this excess:

\begin{itemize}
  \item If the excess is caused by a charge exchange reaction then it should be visible in the stacked spectra of cold clusters, where neutral gas clouds could exist, but not in the stacked spectra of hot clusters.
  \item If the excess is caused by a dark matter decay process, then it should be more significant in the stacked spectra of hot cluster, than in cold clusters.
\end{itemize}

eROSITA (extended ROentgen Survey with an Imaging Telescope Array) is the primary instrument on the Spectrum-Roentgen-Gamma (SRG) mission, which was successfully launched on July 13, 2019 \citep{eROSITA2021}. eROSITA consists of seven individual telescope modules (TMs) arranged in hexagonal shape \citep{eder2018erosita} as shown in Figure \ref{fig:eROSITA2021_TMs}. 

\begin{figure}[ht!]
    \centering
    \includegraphics[width=0.7\textwidth]{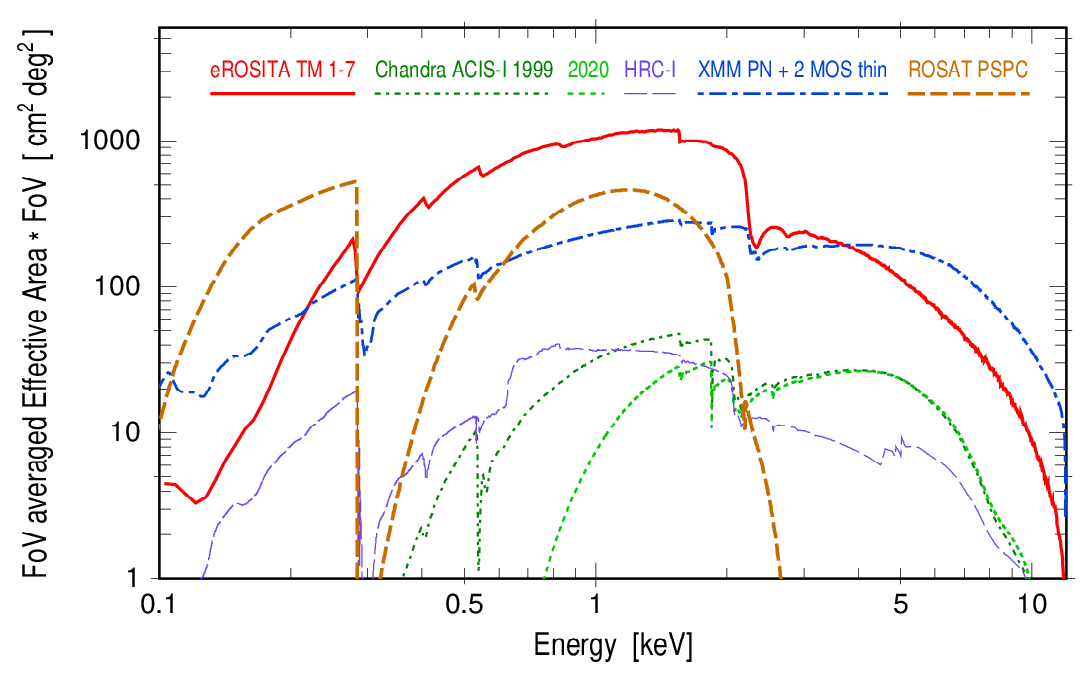}
    \caption { \footnotesize {  Comparison of the product of field of view multiplied by effective area as a function of energy for eROSITA, Chandra, XMM-Newton, and ROSA. Original from \cite{eROSITA2021} }}
     \label{fig:eROSITA2021_grasp}
\end{figure}

Whereas the combined effective area (on-axis) of the seven eROSITA telescopes is lower than that of XMM-Newton pn + MOS in the 3-4 keV band, the product of field of view multiplied by effective area is slightly higher in the 3-4 keV band as shown in Figure \ref{fig:eROSITA2021_grasp}.

For this work we employ the first scan of the eROSITA all-sky survey (eRASS-1). The exposure highly depends on the sky position relative to the ecliptic, ranging from ~100 s at the ecliptic equator to more than ~10,000s close to the ecliptic poles as shown in Figure.

\begin{figure}[ht!]
    \centering
    \includegraphics[width=0.7\textwidth]{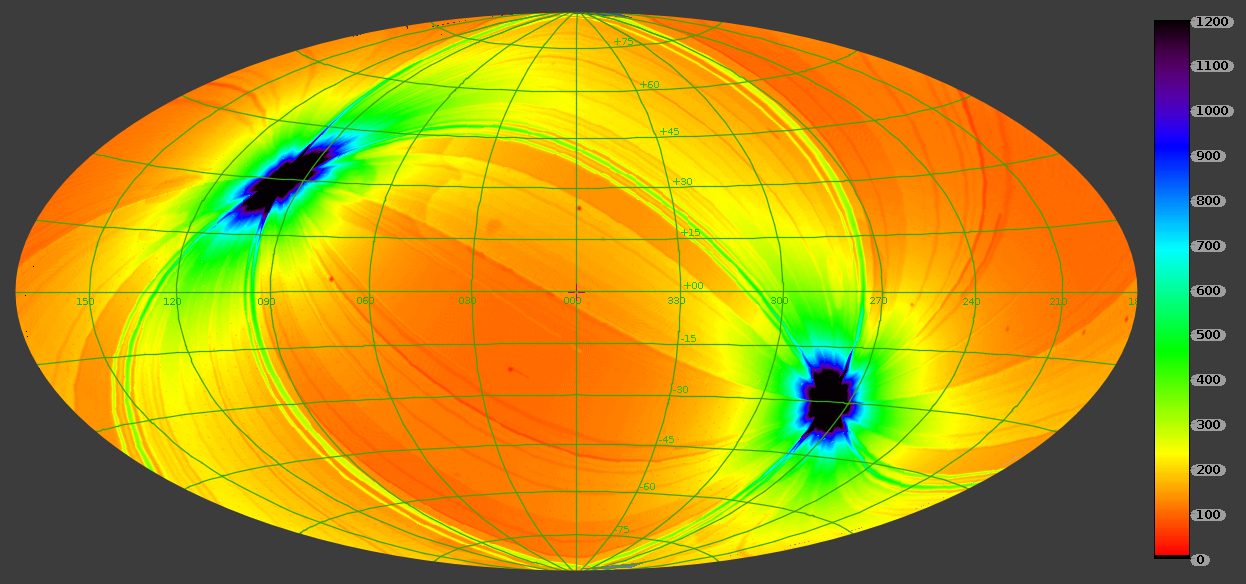}
    \caption { \footnotesize { Effective exposure map in the energy band 0.6–2.3 keV derived for eRASS-1 in galactic coordinates. Original from \cite{eROSITA2021} }} 
     \label{fig:eROSITA2021_exposure}
\end{figure}

Now, given the effective exposure of eRASS-1 ranging from ~100 s to ~10,000s, and if we assume an average exposure of 1000s, we would need about ~8000 clusters to reach the total of 8 Ms that \cite{bulbul2014detection} reached in their combined stacked spectra (2 Ms for PN and 6 Ms for MOS).

%% file: DataProcessing.tex
\chapter{eRASS-1 Cluster data processing} \label{sec:DataProcessing}

\section{Cross-Matching of eRASS-1 with existing catalogs}

\begin{figure}[ht!]
    \centering
    \includegraphics[width=1.0\textwidth]{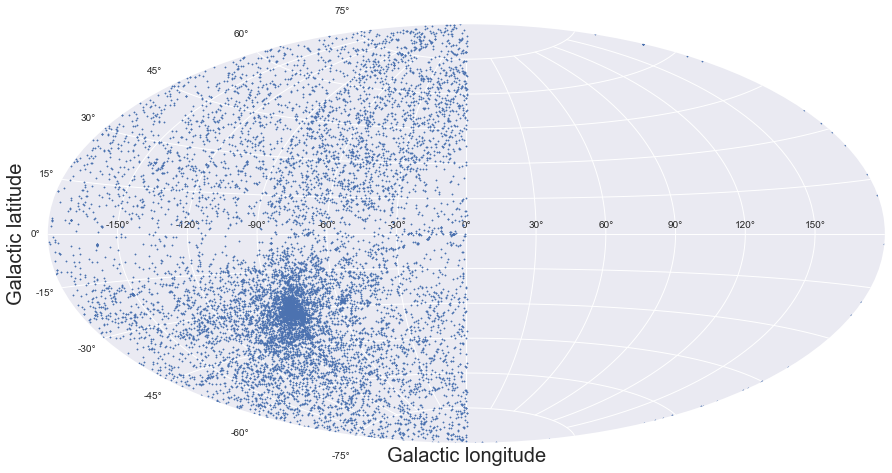}
    \caption { \footnotesize { Clusters detected in eRASS-1 for the German part of the Sky. The extension likelihood threshold to separate point sources from extended sources was set to 6 (EXT\_LIKE \> 6). The accumulation of detections in the lower left side corresponds to the eROSITA poles were the exposure is maximal. } }
     \label{fig:eRASS1}
\end{figure}

The eRASS-1 Clusters and Groups catalogue was created by Ang Liu et all (MPE, Cluster Working Group), using v946 of the eROSITA pipeline, and the eROSITA Science Analysis Software System (eSASS) tasks in the 0.2-2.3 keV band. Further cleaning was applied using the eSASS task FLAREGTI, and the source parameters were obtained with the eSASS task ERMLDET. The extension likelihood threshold to separate point sources from extended sources was set to 6 (EXT\_LIKE \> 6). Figure \ref{fig:eRASS1} shows the resulting 11,028 clusters detected in eRASS-1, in the German part of the eROSITA Sky. 

\begin{figure}[ht!]
    \centering
    \includegraphics[width=0.95\textwidth]{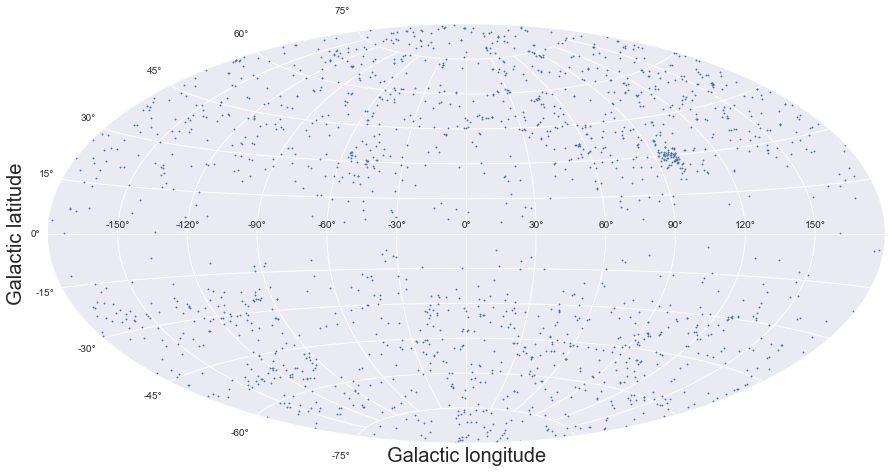}
    \caption { \footnotesize { MCXC: A meta-catalogue of x-ray detected clusters, with a total of 1,743 clusters \citep{MCXC_2011}} }
    \label{fig:MCXC_2011}
\end{figure}

\begin{figure}[ht!]
    \centering
    \includegraphics[width=0.95\textwidth]{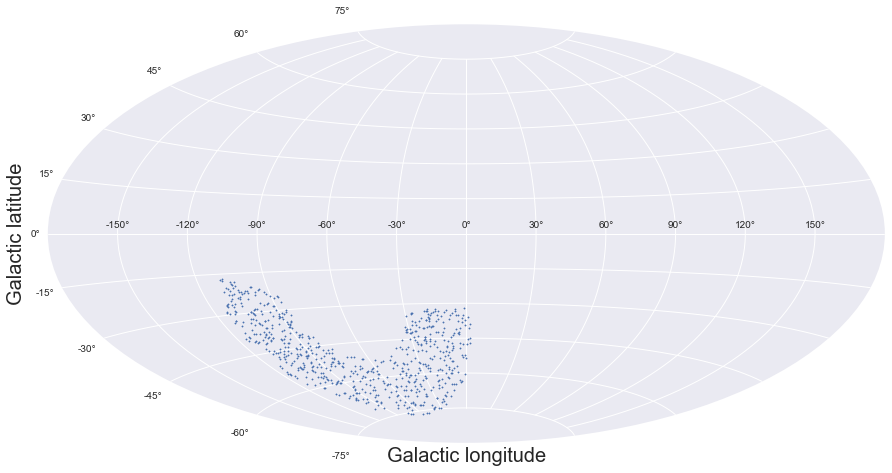}
    \caption { \footnotesize {SPT-SZ 2500d SZ cluster catalog: The South Pole Telescope cluster catalog based on Sunyaev–Zeldovich (SZ) effect, with a total of 677 clusters \citep{SPT_2019}} }
     \label{fig:SPT_2019}
\end{figure}

\begin{figure}[ht!]
    \centering
    \includegraphics[width=0.95\textwidth]{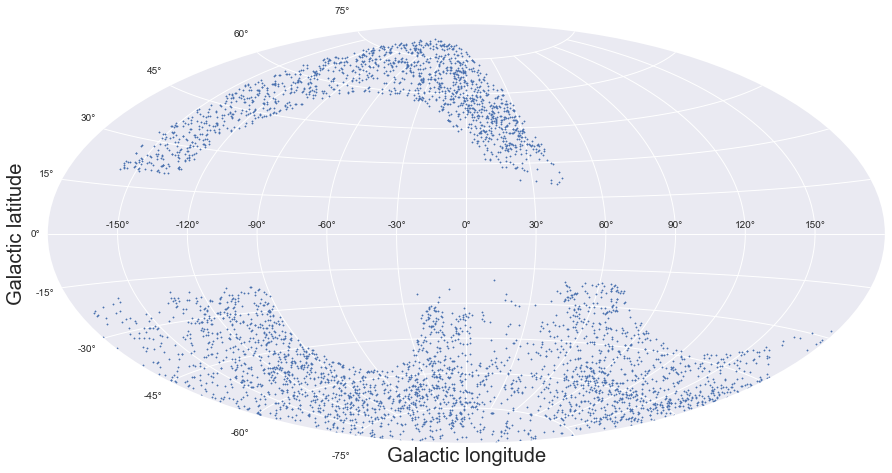}
    \caption { \footnotesize {ACT DR5 Clusters Catalog: The Acatama Cosmology Telescope also based on Sunyaev–Zeldovich (SZ) effect, with a total of 4,195 clusters \citep{ACT_2021}} }
     \label{fig:ACT_2021}
\end{figure}

Notice that the eROSITA point-spread-function (PSF) is quite large, making it difficult to distinguish high-redshift clusters from AGNs as reported by \cite{bulbul2021erosita}. However, in this work we cross-match the clusters listed in the eRASS1 catalog with existing cluster catalogs to limit our sample to confirmed clusters, namely:

\begin{itemize}
  \item The MCXC: A meta-catalogue of x-ray detected clusters \citep{MCXC_2011}, with a total of 1,743 clusters as shown in Figure \ref{fig:MCXC_2011}
  \item SPT-SZ 2500d SZ cluster catalog: The South Pole Telescope cluster catalog based on Sunyaev–Zeldovich (SZ) effect \citep{SPT_2019}, with a total of 677 clusters as shown in Figure \ref{fig:SPT_2019}
  \item ACT DR5 Clusters Catalog: The Acatama Cosmology Telescope also based on Sunyaev–Zeldovich (SZ) effect \citep{ACT_2021}, with a total of 4,195 clusters as shown in Figure \ref{fig:ACT_2021}
\end{itemize}

\begin{figure}[ht!]
    \centering
    \includegraphics[width=0.9\textwidth]{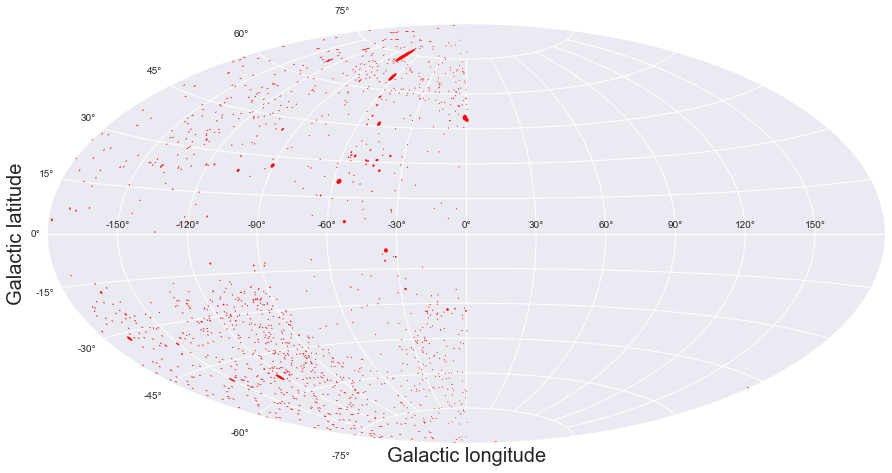}
    \caption { \footnotesize { Resulting sample 1363 clusters after cross-matching the eRASS-1 catalog with the MCXC / SPT / ACT catalogues. The angular size in arcmins is used to show the extend on the sky, with Virgo being the cluster with highest angular size on the sky, close to the Galactic North Pole.} }
    \label{fig:eRASS1_cross_matched}
\end{figure}

For the cross-matching between the eRASS-1 catalog and the MCXC/SPT/ACT catalogues we use a 5-arcmin matching radius, and the duplicates are removed giving first priority to MCXC, second priority to SPT, and third priority to ACT. In this way we obtain a total of 1363 matching clusters. Also, for each matched cluster the R500 radius is transformed from kpc to arcmin by using the WMAP9 package of astropy based on the Nine-Year Wilkinson Microwave Anisotropy Probe (WMAP) Observations \citep{WMAP_2013}. This gives us the sky angular size of the cluster a shown in Figure \ref{fig:eRASS1_cross_matched} which we can use to extract the source and background regions.

\section{Point source masking}

An important step towards obtaining the spectra of each cluster is to remove the contamination by point sources. As described in \cite{bulbul2021erosita}, due to the PSF size of eROSITA, of about 26" on average, point sources such as AGNs may look like extended emission, therefore it is necessary to adapt the radius used to mask point sources depending on the PSF. 

\begin{figure}[ht!]
    \centering
    \includegraphics[width=1.1\textwidth]{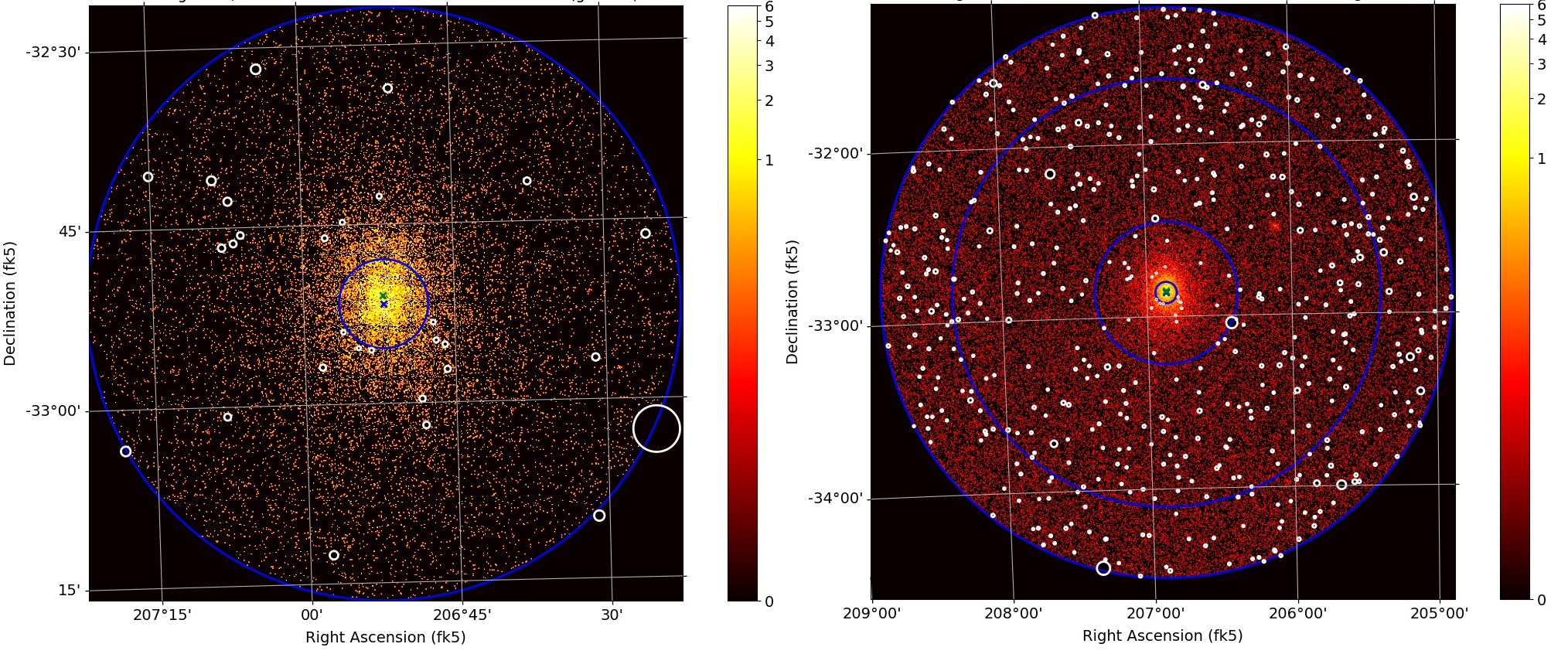}
    \caption { \footnotesize {eRASS-1 image of A3571 after masking the point sources. The left panel zooms in the source region (0-1 R500), and the right panels shows up to 4 times R500 (the outer annuls corresponds 3-4 R500 which is used as background). However notice that we could not mask the point sources in the inner most part of the clusters, corresponding to the [0-0.15] R500 region, since it is not possible to distinguish anymore the extended cluster emission from point sources. The scale indicates total counts. } }
    \label{fig:A3571_mask}
\end{figure}

For this we use a prescription provided by Vittorio Ghirardini (MPE, Cluster Working Group) which finds the radius where the PSF-estimated flux given the count rate (ML\_RATE\_0), exceeds the background (ML\_BKG\_0). This results in an adaptive radius which can mask sources depending on their count rate. Figure \ref{fig:A3571_mask} shows the resulting images of the cluster A3571 after masking the point sources.

\section{$\beta$-model fitting}

The next step involves obtaining a $\beta$-model fit for the source. This helps to identified any un-masked point sources, and also the beta model parameters are necessary to weight the contribution to the effective area when extracting the spectra. 

The $\beta$-model was first proposed by \cite{cavaliere1978distribution}, who considered both an isothermal and an adiabatic equation of state for the gas, where both T and $\sigma_{\rm r} $ are spatially invariant, and dark matter follows the same distribution as galaxies, leading to the relation between the galaxy and gas distributions shown in Equation \ref{eqn:isothermal}

\begin{equation}
    \frac{n_{\mathrm{gas}}(r)}{n_{\mathrm{gas}}(0)}=\left[\frac{\rho_{\mathrm{gal}}(r)}{\rho_{\mathrm{gal}}(0)}\right]^{\beta} \text { with } \beta=\frac{\mu m_{\mathrm{p}} \sigma_{\mathrm{r}}^{2}}{k T}
    \label{eqn:isothermal}
\end{equation}

The full effectiveness of the model arose when \cite{cavaliere1978distribution} proposed to approximate the distribution of Galaxies with the empirical King distribution \cite{king1962structure}, given by Equation \ref{eqn:King2D} for 3D radial distribution, and Equation \ref{eqn:King3D} for 2D projected distribution using the Abell transform \citep{sarazin1986x}:

\begin{equation}
    \left.\rho_{\mathrm{gal 3D}}(r)=\rho_{\mathrm{gal}}(0)\left[1+r / r_{\mathrm{c}}\right)^{2}\right]^{-3/2}
    \label{eqn:King3D}
\end{equation}

\begin{equation}
    \left.\rho_{\mathrm{gal 2D}}(r)=\rho_{\mathrm{gal}}(0)\left[1+r / r_{\mathrm{c}}\right)^{2}\right]^{-1}
    \label{eqn:King2D}
\end{equation}

Replacing Equation \ref{eqn:King2D} into Equation \ref{eqn:isothermal} yields Equation \ref{eqn:beta_model}, the so-called $\beta$-model, where $r_{\rm c}$ is the core radius, and $\beta$ is fixed to $2/3$. For a comprehensive review of the $\beta$-model see \cite{arnaud2009beta}.

\begin{equation}
    n_{\mathrm{gas}}(r)=n_{\mathrm{gas}}(0)\left[1+\left(\frac{r}{r_{\mathrm{c}}}\right)^{2}\right]^{-3 \beta / 2}
    \label{eqn:beta_model}
\end{equation}

The actual fit is done using the astropy Moffat2D model (see Equation \ref{eqn:beta_model}), which is equivalent to the beta model with $r^2  = \left(x-x_{0}\right)^{2}+\left(y-y_{0}\right)^{2}$, $\gamma = r_c$ and fixed $\alpha = 3/2$. Picture x shows the the fitted beta model for A3571, in particular the right side panel shows the residuals where no remaining unmasked point source is outlaying, but it is possible to appreciate some structure, probably due to AGN emission.

\begin{figure}[ht!]
    \centering
    \includegraphics[width=1.1\textwidth]{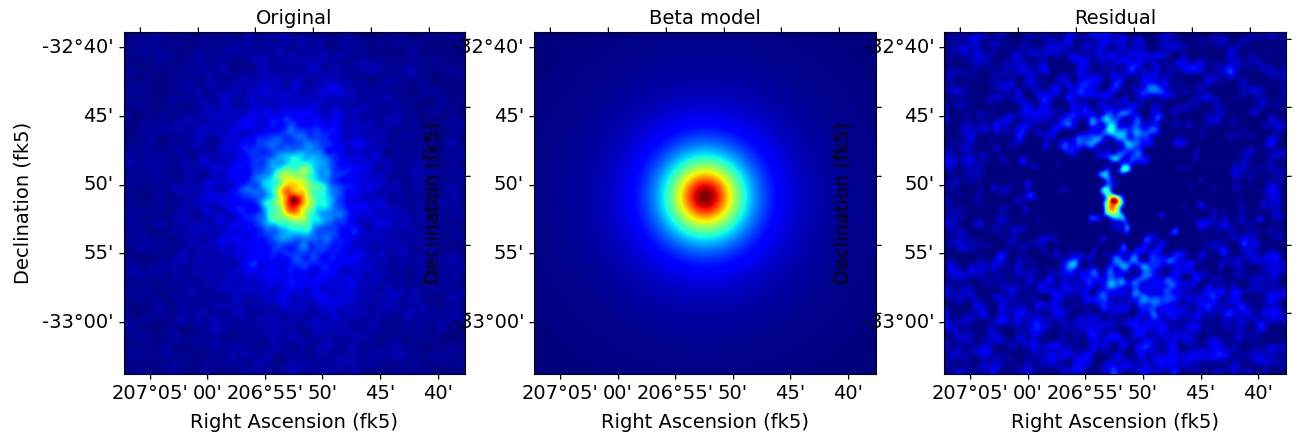}
    \caption { \footnotesize {Left: eRASS-1 image of A3571 after applying a $1\sigma$ Gaussian smoothing filter to the image. Middle: $\beta$-model fitted to the Gaussian smoothed image, Left: Residual after removing the fitted $\beta$-model from the Gaussian smoothed image. The scale indicates total counts. } }
    \label{fig:A3571_beta}
\end{figure}

\section{Spectra extraction and fitting}

Once the contaminating point sources have been masked, and the source model parameters retrieved, the next step is to extract the spectra with the eSSAS task srctool. The extraction region for the source spans [0-1] R500, and for the background [3-4] R500. At this stage it is important to notice that the time and spacial resolution play a key role in the processing time of the spectra extraction, so a compromise is needed:

\begin{itemize}
    \item Time accuracy parameter (tstep [s]): This parameter controls the time resolution in the tracking of the spacecraft pointing, it affects the calculation of the source coordinates and extraction regions on the detector focal plane. The processing time scales approximately as a function of $tstep^{-1}$
    \item Spatial accuracy parameter (xgrid): This parameters controls the linear spatial sampling used to calculate the fractional response in units of the eROSITA physical pixel scale. The processing time scales approximately as a function of $xgrid^{-2}$.
\end{itemize}

Since the processing time scales linearly for the time resolution, but quadratic for the spatial resolution, we decided to use the native eROSITA physical pixel scale of ~8arcsec (xgrid = 1.0), but an increased time accuracy parameter to have finer resolution, of tstep = 0.3s. This is aligned with the values commonly used for the eROSITA Final Equatorial-Depth Survey (eFEDS) X-Ray analysis (see \cite{liu2021erosita} and \cite{bahar2021erosita} for example).

\begin{figure}[ht!]
    \centering
    \includegraphics[width=0.7\textwidth]{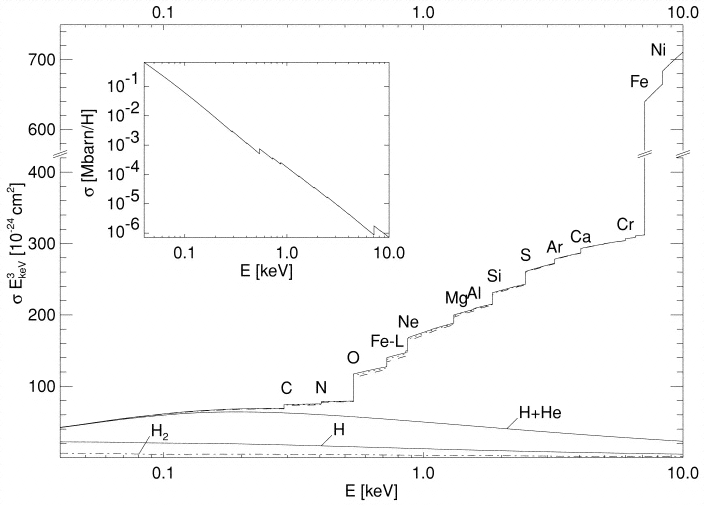}
    \caption { \footnotesize {Absorptivity per hydrogen atom of the ISM, the inset shows the cross section without the multiplication by $E^3$. The dotted line shows a model using a MRN distribution for the grains, whereas the dashed line assumes that all grains have a radius of $a = 0.3 \mu m$. Original from \cite{wilms2000absorption} } }
    \label{fig:tbabs}
\end{figure}

Once the spectra has been extracted the next step is fitting it using the APEC/ADED model described in Section \ref{sec:APEC}. Additionally it is necessary to take into account the absorption by the Inter Stellar Medium (ISM) as shown in Figure \ref{fig:tbabs}. For this we use the Tuebingen-Boulder ISM absorption model \citep{wilms2000absorption}. This model takes into account gas, grains and molecules. The different elements in the gas phase are weighted by abundance, taking into account depletion onto grains. However we use the photoelectric absorption cross sections with variable abundances from \citep{balucinska1992photoelectric}.

Both the APEC/APED model for the X-Ray emission of the Intra Cluster Medium (ICM), and the X-Ray absorption of the Inter Stellar Medium (ISM) are implemented in the Xspec X-Ray Analysis Package \citep{arnaud1996xspec}. In the xspec terminology we adopt the tbabs*apec model, where the composed model parameters are as follows:

\begin{itemize}
    \item Galactic Hydrogen Column: This is the main parameter of the absorption model, for this we adopt the values from \cite{willingale2013calibration}, which can be retrieved online using the central coordinates of the cluster. It is important to consider that for close clusters, with large angular extension on the sky, the hydrogen column can vary across the angular extension of the cluster. In this cases the hydrogen column can be let free in the model, to obtain an average effective hydrogen column.
    \item Temperature: Plasma temperature assuming a hot, optically thin plasma that is in collisional ionization equilibrium. It determines the spectral slope of the bremsstrahlung emission, the ionization balance of the various elements, and therefore the ratio of the line emissions.
    \item Metal abundance: Metal abundances with respect to solar. For the [0-1] R500 region of a cluster it is typically close to 0.3, but it can be higher for smaller regions close to the core of the cluster. For this work we use the abundance tables from \cite{asplund2009chemical}.
    \item Norm: The model norm depends determines the total X-Ray flux, which depends on primarily on the electron density of the cluster but also on the redshift and cosmological model used. For this work we use $\Omega_M = 0.3$ and $\Omega_{\lambda}=0.7$
\end{itemize}

Also it is important to notice that for the individual clusters the number of counts is not very high, so it is necessary to use Poisson statistics to properly subtract the background, with maximum likelihood-based statistic  given by \cite{cash1979parameter} and shown in Equation \ref{eqn:cstat}, where $S_i$ are the observed counts, $t$ the exposure time, and $m_i$ the predicted count rates based on the current model and instrumental response:

\begin{equation}
    C=2 \sum_{i=1}^{N}\left(t m_{i}\right)-S_{i}+S_{i}\left(\ln \left(S_{i}\right)-\ln \left(t m_{i}\right)\right)
    \label{eqn:cstat}
\end{equation}

Even comprehensive comparisons of Gaussian versus Poisson statistics have shown that for a broad class of problems, unless the number of data bins is far smaller than $\sqrt{N_c}$, where $N_c$ is the total number of counts in the dataset, the bias will still likely be comparable to, or even exceed, the statistical error \cite{humphrey2009chi2}.

\begin{figure}[ht!]
    \centering
    \includegraphics[width=1.0\textwidth]{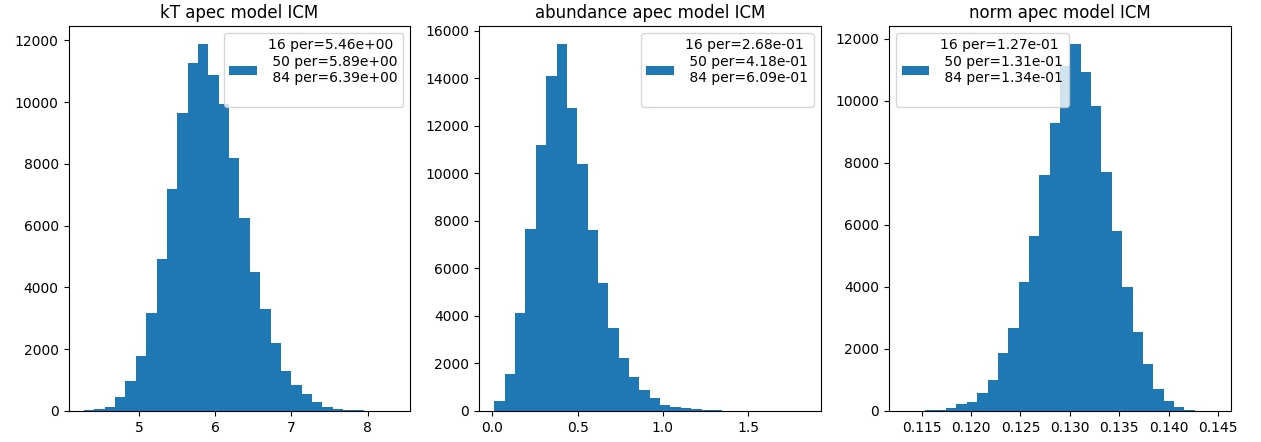}
    \caption { \footnotesize {Resulting Markov Chain distributions for the temperature, norm and abundance of A3571. } }
    \label{fig:A3571_mcmc}
\end{figure}

\begin{figure}[ht!]
    \centering
    \includegraphics[width=1.0\textwidth]{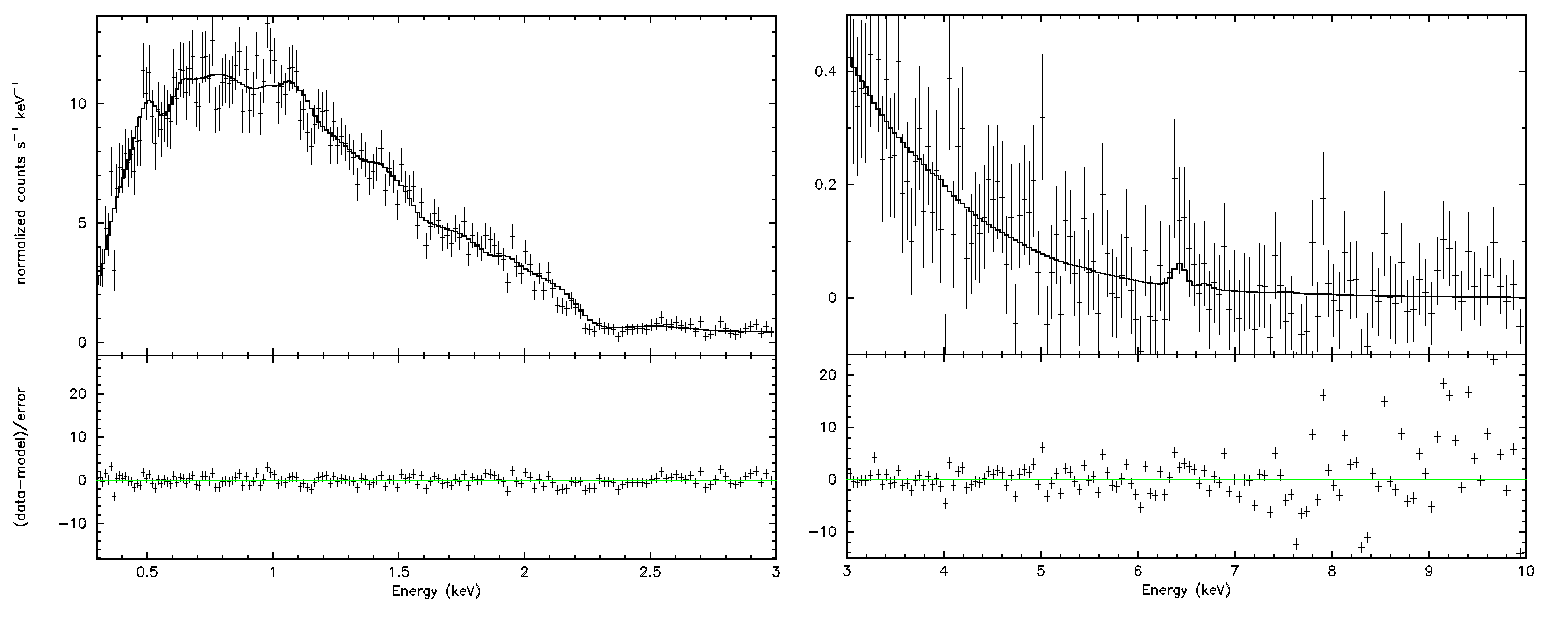}
    \caption { \footnotesize { Spectral fit for A3571 obtained by setting the free parameters (temperature, abundance and norm) with the mean values of the Markov Chain distributions. } }
    \label{fig:A3571_spectra}
\end{figure}

As fitting algorithm we use a Markov Chain, namely Goodman-Weare algorithm, with a burn phase of 1000 steps, and a length of 10000 steps. As a prior we use the Gaussian distribution, with covariances taken from a pre-fit done using the Levenberg-Marquardt algorithm based on the CURFIT routine from Bevington. An example of the resulting Markov Chain distributions for the temperature, norm and abundance of A3571 is shown in Figure \ref{fig:A3571_mcmc}, together with the corresponding spectral fit, obtained by setting the free parameters with the mean values of the Markov Chain distributions, shown in Figure \ref{fig:A3571_spectra}.

\section{Cluster sample statistics}

\begin{figure}[ht!]
    \centering
    \includegraphics[width=1.0\textwidth]{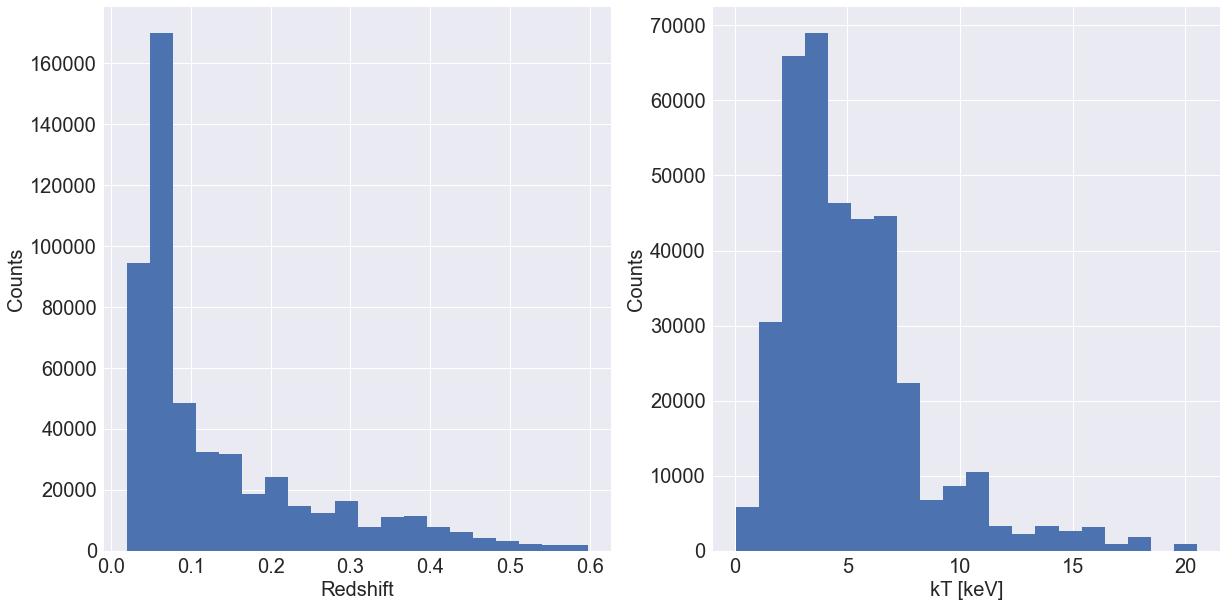}
    \caption { \footnotesize { Count distributions for redshift (left panel) and temperature (right panel) } }
    \label{fig:kT-z-hist}
\end{figure}

\begin{figure}[ht!]
    \centering
    \includegraphics[width=1.0\textwidth]{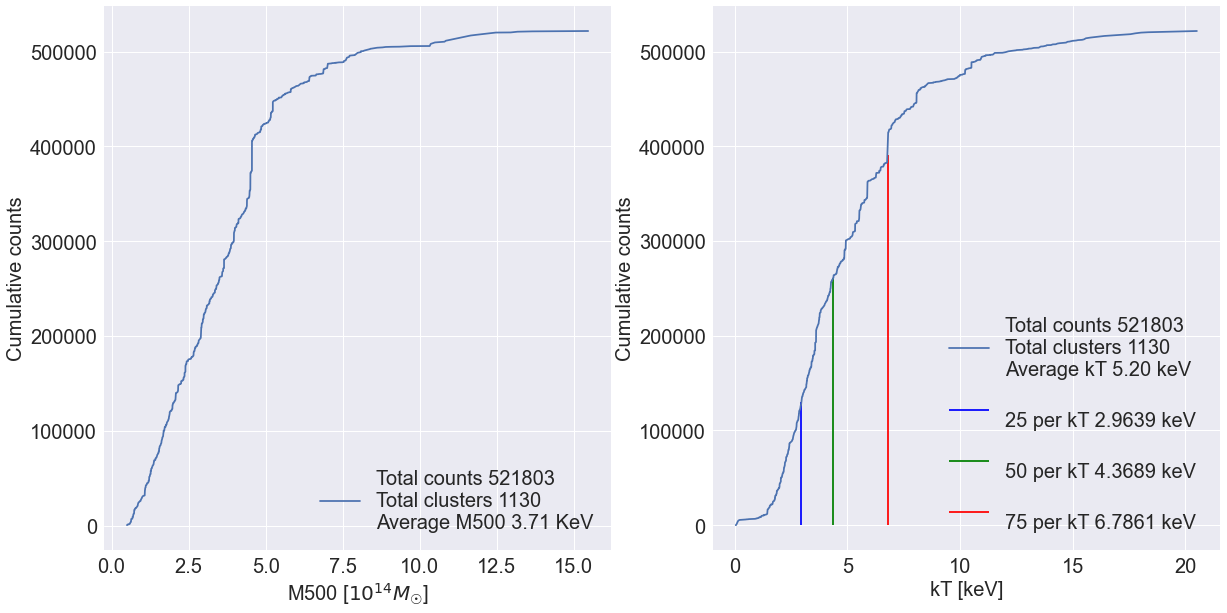}
    \caption { \footnotesize { Cumulative count distribution for redshift (left panel) and temperature (right panel), indicating the 3 temperatures that define 4 bins with equal amount of counts. } }
    \label{fig:kT-z-cum}
\end{figure}

Once we have the temperatures of all clusters we can group them in 4 separated temperature bins, each with equal amount of total counts. Figure \ref{fig:kT-z-hist} shows the count distribution for redshift and temperature, and Figure \ref{fig:kT-z-cum} the cumulative count distribution, indicating the 3 temperatures that define 4 bins with equal amount of counts: $<2.96keV$, $[2.96-4.37]keV$, $[4.37-6.79]keV$, $>6.79keV$. Also the median temperature (4.37keV) can be used to define two groups of temperature to increase the counts in each group, and thus the signal to noise (SNR) ratio.

%% file: Calibration.tex
\chapter{eROSITA calibration status} \label{sec:calibration}

\section{TM5 and TM7 light leak} \label{sec:lightLeak}

\begin{figure}[ht!]
    \centering
    \includegraphics[width=0.9\textwidth]{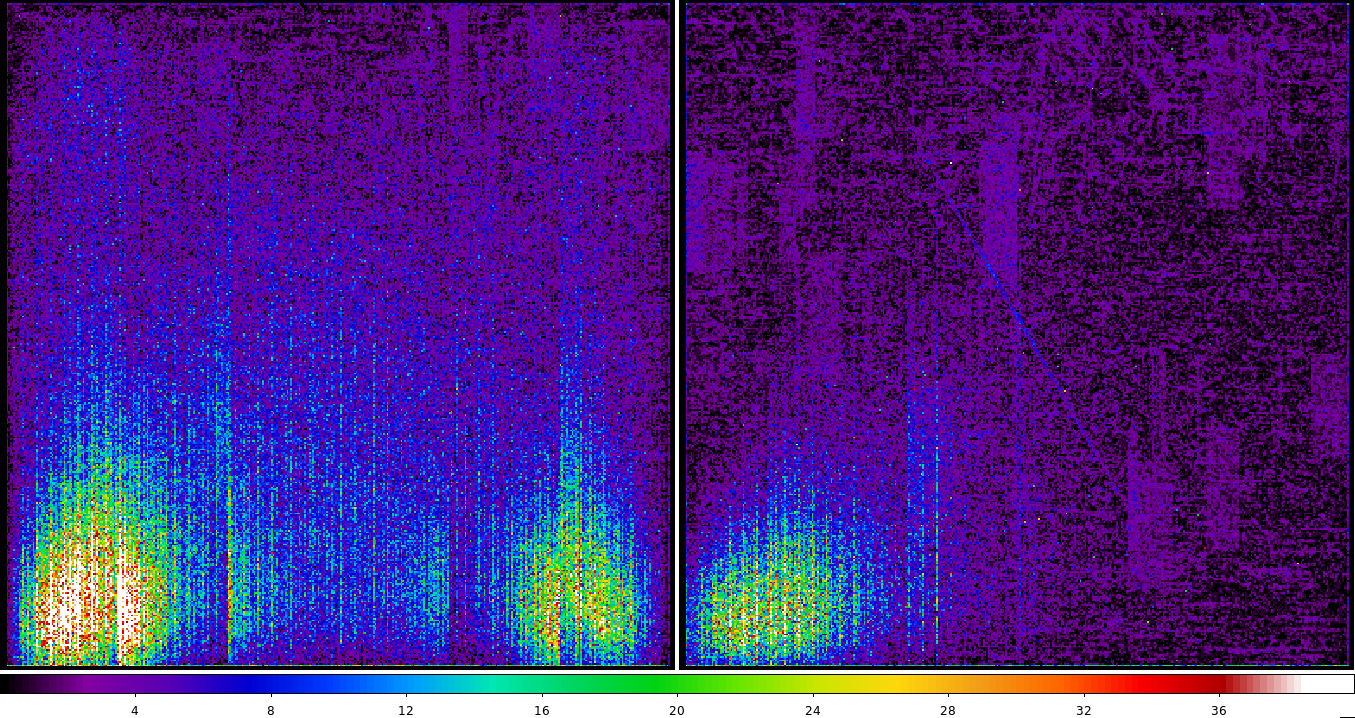}
    \caption { \footnotesize { Images of the CCD cameras of TM5 (left) and TM7 (right) showing the contamination caused by the optical light leak. Original from eROSITA-DE: Early Data Release site} }
     \label{fig:TM57_lightleak}
\end{figure}

During the commissioning phase of eROSITA it was noticed that two of the telescope modules, namely TM5 and TM7 had optical light contamination in the lower part of their respective CCDs as shown in Figure \ref{fig:TM57_lightleak}. This problem only affects TM5 and TM7 because there was a plan two use these TMs for low-energy spectroscopy, and therefore they do not have an aluminium on-chip optical light filter like the other 5 cameras as describe in \cite{eROSITA2021}.

\begin{figure}[ht!]
    \centering
    \includegraphics[width=1.0\textwidth]{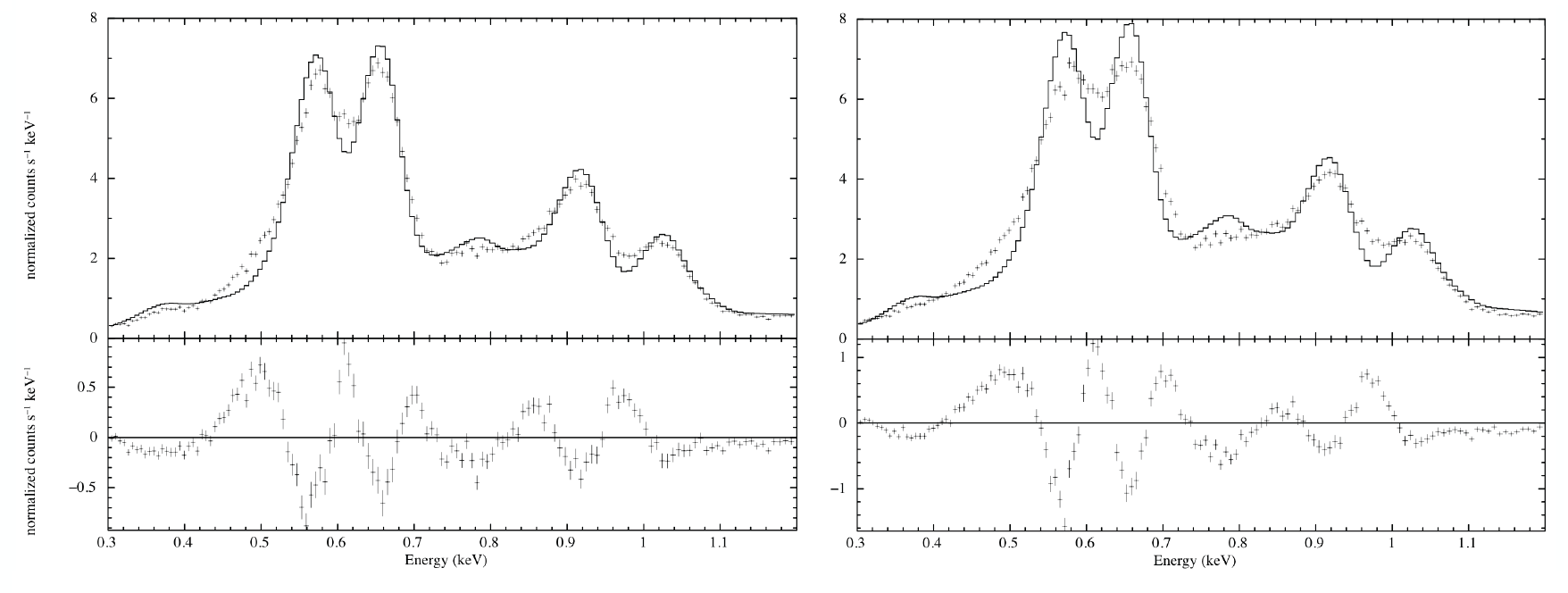}
    \caption { \footnotesize { TM5 (left) and TM7 (right) spectral fit and residuals for the Super Nova Remnant 1E 0102.2-7219, a typical target used for cross-calibration as described in \citep{plucinsky2012cross}. Original from Paul P. Plucinsky (Smithsonian Astrophysical Observatory) 2021/06/17 } }
     \label{fig:TM57_SNR_fit}
\end{figure}

This contamination known as "light leak" comes from the Sun, and is mostly limited below 0.8keV, however it impacts the energy scale calibration over the whole energy range of the CCDs, resulting in important biases when fitting the spectra as shown in \ref{fig:TM57_SNR_fit}, corresponding to the Super Nova Remnant 1E 0102.2-7219, a typical target used for cross-calibration as described in \citep{plucinsky2012cross}.

Since the light contamination comes from the Sun, it highly depends on the orientation of the spacecraft, which varies throughout one complete survey. However, at the present moment the Sun angle constraints are not yet defined. Therefore this work makes no use of TM5 and TM7 data, given that that the light leak affects significantly the precision of the  spectral analysis, which is critical when trying to detect faint signal based on fit residuals.

\section{ A3266 Calibration target residuals} \label{sec:A3266}

\begin{figure}[ht!]
    \centering
    \includegraphics[width=1.0\textwidth]{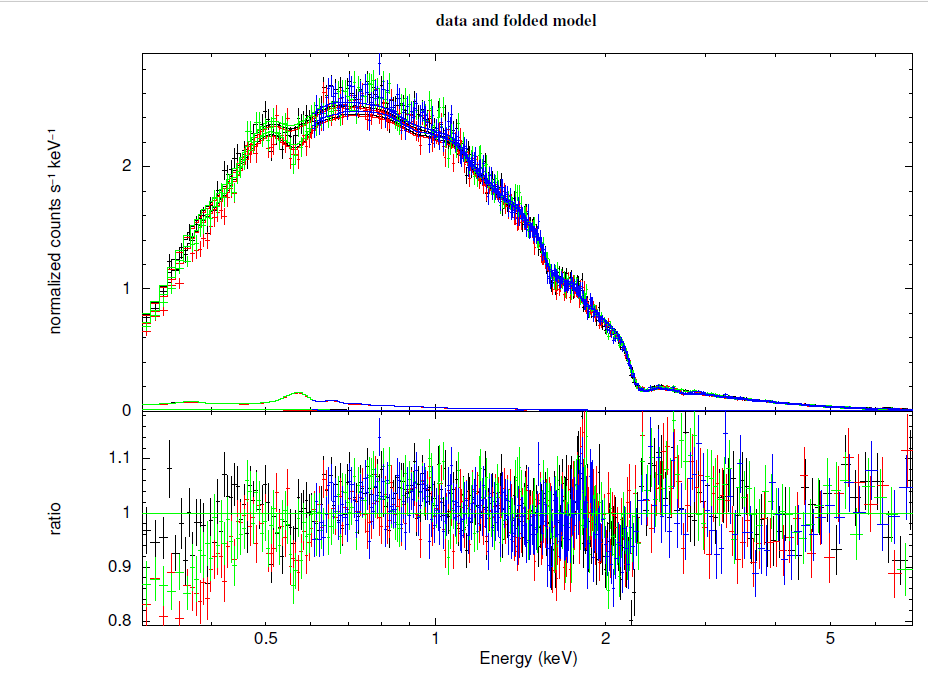}
    \caption { \footnotesize { Fit to the spectra of A3266 extracted from the inner 6arcmin region. Residuals at the 10\% are present, particularly at the edges of the effective area curves. Original from \cite{sanders2021studying} } }
     \label{fig:A3266}
\end{figure}

Also, and as reported by \cite{sanders2021studying}, the spectral fit for the eROSITA calibration target A3266 shows residuals at the 10\% level for the entire energy range, particularly at the edges of the effective area curves as shown in Figure \ref{fig:A3266}. These residuals are present even if TM5 and TM7 are excluded, so the problem it is not related to the light leak. At the present moment the origin of these residuals has not been clarified, and therefore it affects the analysis presented in this work.

\section{Gold edge in the vignetting function} \label{sec:vignetting}

In this work we have considered the possibility, that the fit residuals seen for the calibration target A3266 could be related with a sparse sampling of the vignetting correction, which does not sample the gold edge. As described by \cite{dennerl2020calibration} the vignetting function has been measured in the PANTER X-Ray test facility at some specific energies corresponding to absorption features (C-K at 0.28 keV, Cu-L at 0.93 keV, Al-K at 1.49keV, Ag-L at 2.50keV, Ti-K at 4.51keV, Cr-K at 5.41keV, Fe-K at 6.40keV, Cu-K$\alpha$ at 8.04keV and Ge-K$\alpha$ at 9,88keV). An  analytical function is used to interpolate in between these energies, but the gold edge in between Al-K (1.49 keV) and Ag-L (2.98 keV) is not well modelled by the interpolation function as it is a highly non-linear feature. Actually ray-tracing simulations from Peter Friedrich (MPE) show a fast, non linear behaviour of the vignetting function at the gold edge, as seen in Figure \ref{fig:A3266}.

\begin{figure}[ht!]
    \centering
    \includegraphics[width=1.0\textwidth]{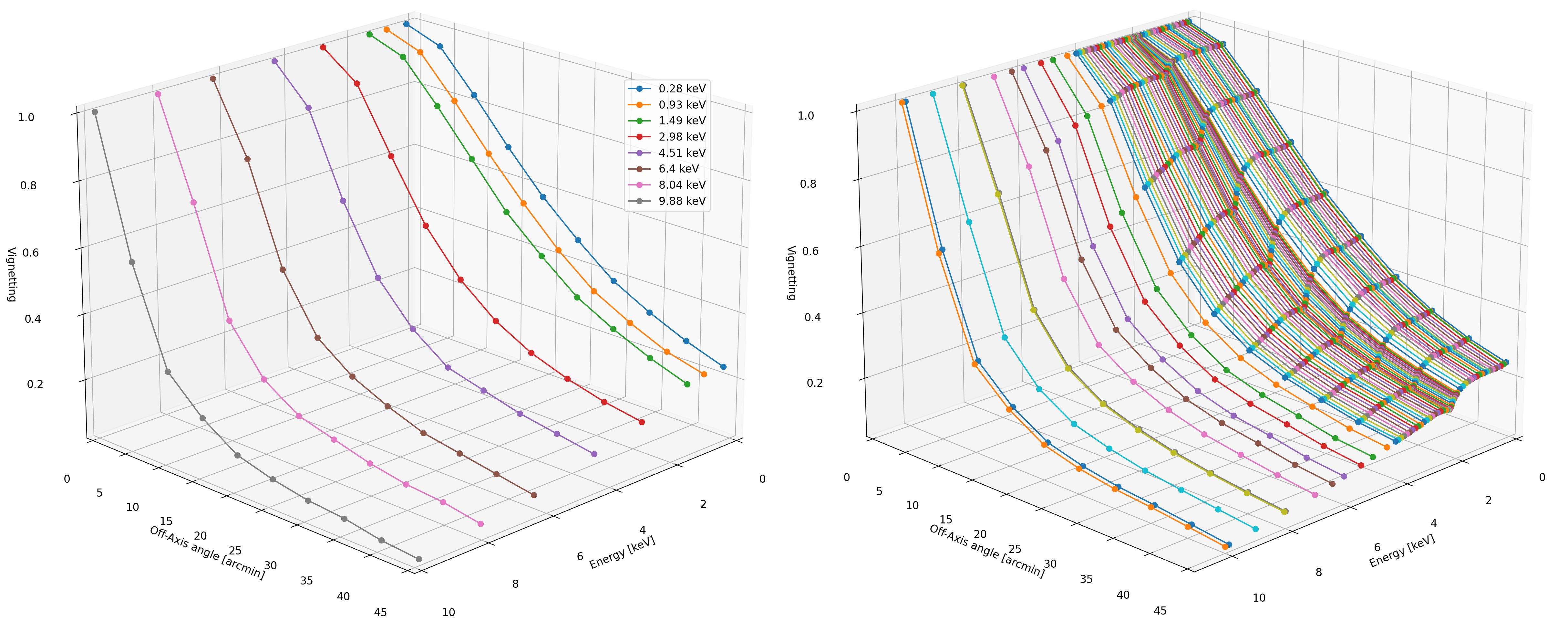}
    \caption { \footnotesize { Left: Vignetting function measure for TM1 at the line energies, which do not sample the gold edge. Right: Vignetting function obtained from ray-tracing simulations, including finer 0.02keV sampling at the gold edge in between Al-K (1.49 keV) and Ag-L (2.98 keV).  } }
     \label{fig:vignetting}
\end{figure}

\begin{figure}[ht!]
    \centering
    \includegraphics[width=1.0\textwidth]{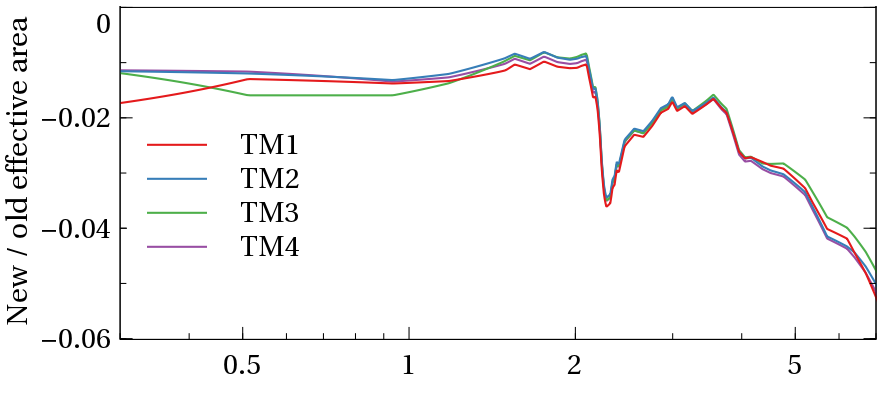}
    \caption { \footnotesize { Ratio of effective areas for A3266 with and without including the gold edge in the vignetting function. Original from Jeremy S. Sanders 2021/11/23 } }
     \label{fig:area_ratio}
\end{figure}

However, for the case of A3266, described in the previous section, the inclusion of the gold edge in the vignetting function changes the effective area only by a maximum of only 4\% at the gold edge as seen in Figure \ref{fig:area_ratio}. Therefore the presence of the gold edge in the vignetting function, cannot explain alone the residuals at the 10\% level, which are seen for the entire energy range.

%% file: Stack.tex
\chapter{Shifting and Stacking technique} \label{sec:Shifting and stacking technique}

\section{Concept} \label{sec:concept}

\begin{figure}[ht!]
    \centering
    \includegraphics[width=0.7\textwidth]{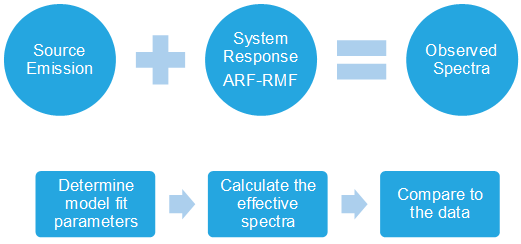}
    \caption { \footnotesize { Process to obtain the actual predicted spectrum,so that it can be compared with the data.} }
     \label{fig:ARF-RMF}
\end{figure}

This work is based on the stacking procedure outlined in \cite{bulbul2014detection}. It requires shifting of the spectra, auxiliary response files (ARF) and redistribution matrix files (RMF), which are necessary to fold the model with the system response to obtain the actual predicted spectrum, as shown in Diagram \ref{fig:ARF-RMF}. In particular the ARF file determines the effective area, also shown in Figure \ref{fig:eROSITA2021_grasp}, and the RMF file determines how incident photons are redistributed among channels. Both ARF and RMF file depend on the frequency (energy), so it is necessary to shift them so that they are aligned with the event files.

After shifting all individual spectra and ancillary files, the stacking procedure takes care of combining the source and background spectra. This requires accounting for the exposure and background scaling, as well as obtaining average ARF and RMF files, weighted by the individual X-Ray flux of each contributing spectra. The stacked spectra can still can be modelled with the same plasma codes such as APEC/APED \citep{smith2001collisional}, by properly accounting for multi-temperature distribution.

\section{Spectra Shifting and Randomization} \label{sec:spectra}

Following \cite{bulbul2014detection}, to shift the spectra we multiply the energies of the photon events by 1+redshift, in order to undo the red-shifting caused by the cosmological expansion, and change the spectra to the source frame as shown in Equation \ref{eqn:shift}. This process aligns the spectra of different clusters at different red-shifts, so that they can be stacked in the same reference frame. Notice that it is necessary to shift directly the energies in the event files, because the conversion from photon events to counts takes as input the event files, and also because it is necessary to apply randomization after shifting.

\begin{equation}
\begin{aligned}
E_{\mathrm{\ source}} = E_{\mathrm{\ observed}} \cdot \left( 1 + \mathrm{redshift} \right)
\end{aligned}
\label{eqn:shift}
\end{equation}

\begin{figure}[ht!]
    \centering
    \includegraphics[width=0.9\textwidth]{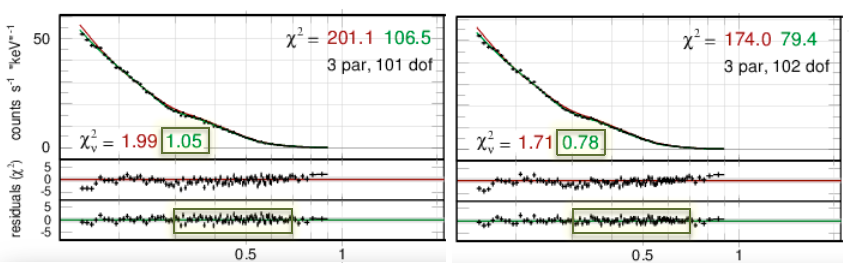}
    \caption { \footnotesize { 2 different randomization repetitions for the XMM-Newton spectra of the Neutron Start RXJ 1856. Both cases employ the same original event files, and models, and the only difference comes from values drawn of the uniform distributions used for randomization. Green indicates the lower end of the $\chi^2$ distribution, and color indicates the higher end of the $\chi^2$ distribution. Notice how the scatter and outliers significantly change from comparing both repetitions. Original from \cite{dennerl2017randomization} } }
     \label{fig:randomization_Konrad}
\end{figure}

Randomization is a process which consist of re-assigning the energy of each photon event, by drawing the new energy value from an uniform distribution with average value corresponding to the original energy as shown by Equation \ref{eqn:randomization}. \cite{bulbul2014detection} used as width of the distribution that of the matrix re-distribution file (RMF).  This process prevents truncation/decimation effects, since the energy column of the event files typically has integer format. On the other hand, and as described by \cite{dennerl2017randomization} randomization introduces fluctuations, increases the $\chi^2$ residuals, and complicates the reproducibility of the results as shown in Figure \ref{fig:randomization_Konrad}, therefore it is desirable to minimize the need for randomization.

\begin{equation}
\begin{aligned}
E_{\mathrm{\ randomized}} = E_{\mathrm{\ original}} \cdot \left[ \mathrm{random.uniform}\left( -\frac{\mathrm{Width}}{2},+\frac{\mathrm{Width}}{2} \right) \right]
\end{aligned}
\label{eqn:randomization}
\end{equation}

 In this sense \cite{dennerl2017randomization} has developed a novel concept for eROSITA, where the energy column of the events files (PI column) is stored as a float instead of integer to minimize truncation/decimation effects, and is computed from integer raw amplitude values (PHA column) by multiplication with real valued gain and charge transfer inefficiency (CTI) correction factors. As these are energy dependent and differ from pixel to pixel, it is expected that the real valued PI energy column exhibits a distribution which is smooth enough that the need for randomization is minimized.
 
In general, and also as outlined by \cite{dennerl2017randomization}, the correct approach to apply randomization would be to propagate the probability distribution of each real PHA value within the PHA bin through the whole processing pipeline. But for this work, we have tested in practice a direct randomization of the PI column over the interval which corresponds to the width of the transmitted PHA bin. For energies above ~1 keV the PHA bin width is ~0.8 eV, below it raises to ~1.1 eV in a more complicated way, depending on the apply on-board binning to save telemetry, which increases the effective bin size as shown in Table \ref{tab:pha_bin}

\begin{table}[ht!]
  \centering
    \begin{tabular}{||c c||} 
        \hline
        Energy Range [keV] & Randomization width [eV] \\ 
        \hline\hline
        0.3-1.0 & 1.1 \\ 
        \hline
        1.0-2.8 & 0.8 \\
        \hline
        2.8-8.63 & 0.4 \\
        \hline
    \end{tabular}
  \caption{ \footnotesize { Effective randomization width taking into account on-board binning of PHA values}}
  \label{tab:pha_bin}
\end{table}

\begin{figure}[ht!]
    \centering
    \includegraphics[width=1.1\textwidth]{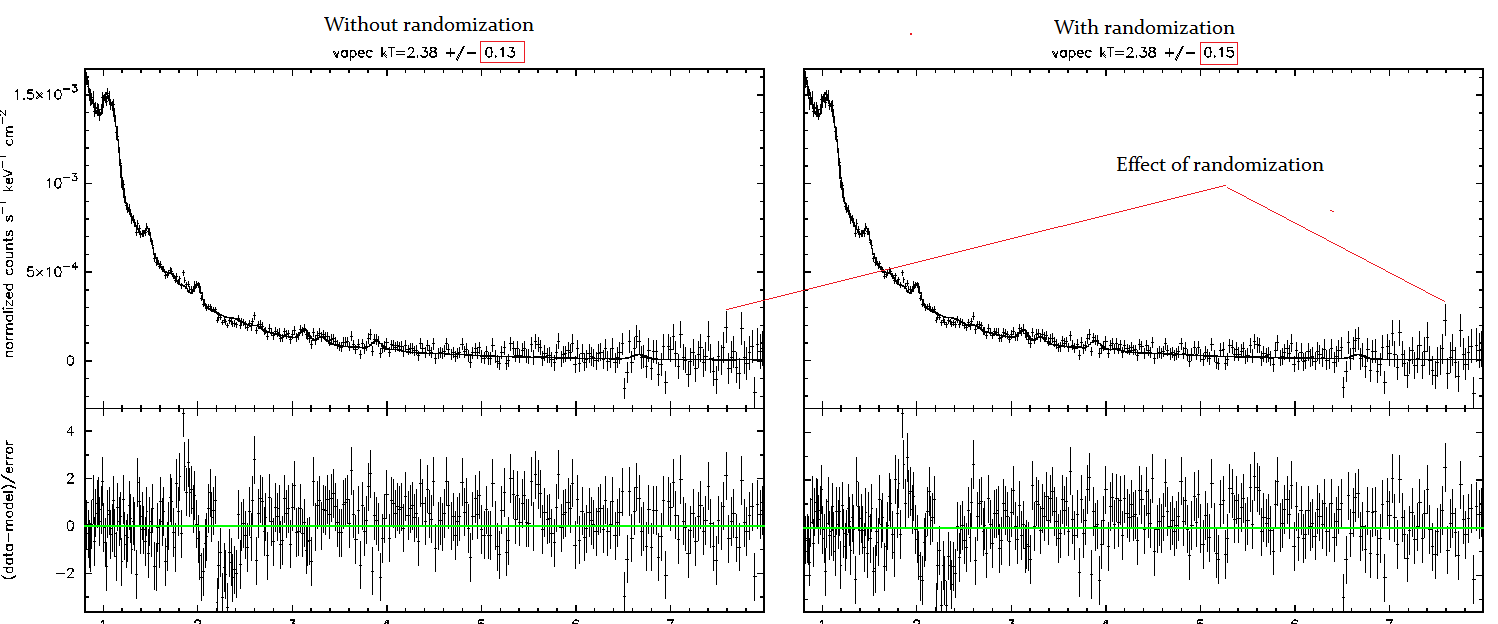}
    \caption { \footnotesize { Effect of the randomization schema based on PHA binning and applied to the cold half of the cluster sample (left: without randomization, right: with randomization) } }
     \label{fig:randomization_Justo}
\end{figure}

As a test we have applied this randomization schema to the stacked spectra of the cold half of the cluster sample. The results are shown in Figure \ref{fig:randomization_Justo}. Randomization barely changes the resulting spectra and is only noticeable at the very high energies ($>7keV$) where the effective area and counts are very small, however the randomized version has higher error in the fitted temperature.

\section{ARF Shifting} \label{sec:ARF}

ARF shifting consist of finding the effective area values which correspond to the shifted spectra. In the case of \cite{bulbul2014detection} this was done by finding the nearest energy bin, corresponding to the un-shifted energies, and assigning the corresponding effective area value to the shifted energy values. The energy columns themselves (ENERG\_LO, ENERG\_HI), remain unchanged, and only the effective area column (SPECRESP) is modified. 

\begin{figure}[ht!]
    \centering
    \includegraphics[width=1.0\textwidth]{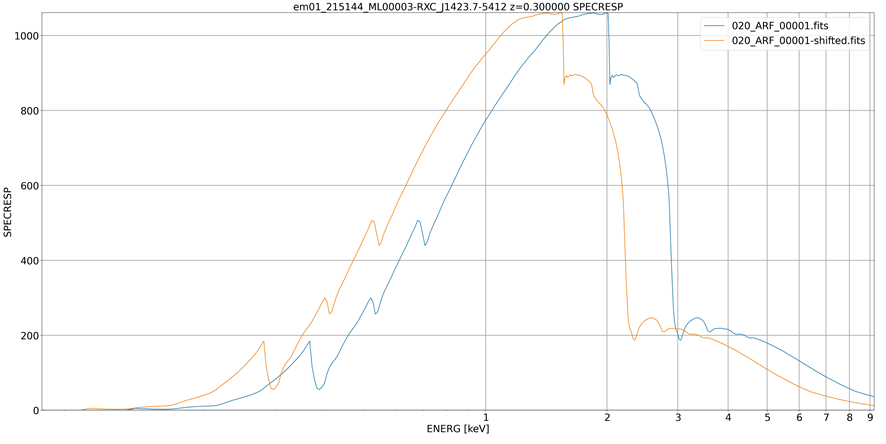}
    \caption { \footnotesize { Example of effective area shifting for RXC J1423.7-5412, a high red-shift cluster at z=0.3. Notice how the fine structure at the gold edge is nicely preserved by the cubic interpolation algorithm. } }
     \label{fig:ARF-shifting}
\end{figure}

For this work, we have used interpolation to find the effective area values corresponding to the shifted energies. The interpolation algorithm is cubic, with implementation from the python package scipy.interpol. This approach allows to preserve the fine structure of the ARF as shown in Figure \ref{fig:ARF-shifting}.

\section{RMF Shifting} \label{sec:RMF}

RMF shifting is a more complicated process, since the redistribution matrix is a two dimensional structure, along photon energy, and channels energy bins. The RMF distribution describes the probability that an incident photon of a given energy $E_{\mathrm{photon}}$, is detected at a given channel bin $ E_{\mathrm{photon}}$, as shown in Equation \ref{eqn:RMF-distribution}, where $\sigma$ relates to the spectral resolution that depends on the energy of the incident photon.

\begin{equation}
\begin{aligned}
\mathrm{RMF} \left( E_{\mathrm{channel}}, E_{\mathrm{photon}} \right) =\frac{\mathrm{Norm}}{\sigma \sqrt{2 \pi}} e^{-\frac{ (E_{\mathrm{channel}} - E_{\mathrm{photon}} )^{2} }{2 \sigma^{2}}}
\end{aligned}
\label{eqn:RMF-distribution}
\end{equation}

\cite{bulbul2014detection} applied two different RMF shifting schemes, one for MOS and another for PN:

\begin{itemize}
    \item Shifting of RMF corresponding to MOS data: This was a custom two-dimensional nearest neighbour algorithm, designed to find the re-distribution probability corresponding to a given (photon energy, channel bin) pair in the un-shifted frame, and assigning the value to the shifted (photon energy, channel bin) pair. The photon energy, and channel bin columns (ENERG\_LO , ENERG\_HI, E\_MIN, E\_MAX) remain unchanged.
    \item Shifting of RMF corresponding to PN data: This was done using ftgcorrmf, a tool of the NASA's HEASARC Software package for space X-Ray missions \citep{drake2016heasarc}. In practice a developer version of this tool was used, which allows to simultaneously shift energy and channels, however the current published version only supports shifting along either energy or channels, but not simultaneously, nor in a two-step process.
\end{itemize}

For this work we contacted HESARC team, to know if ftgcorrmf could be used to shift both energy and channels for eROSITA data. Keith Arnaud confirmed that the current version of ftgcorrmf is designed to shift either channels or energy, not both. Therefore we decided to follow the RMF shifting procedure that \cite{bulbul2014detection} used for MOS data, but rather than applying a two-dimensional nearest neighbour algorithm, we apply a  two-dimensional cubic interpolation algorithm, using the python package scipy.interpol.

It is important to notice that the RMF shifting process does not only shift the distribution, but also also widens it, and therefore must be re-normalized:
 
\begin{itemize}

    \item The peak of the distribution in the channel bin space is centered at the energy of the incident photon, and after shifting it is centered at the shifted energy as shown by Equation \ref{eqn:RMF-shifting}:
    
\begin{equation}
\begin{aligned}
\mathrm{RMF} \left( E_{\mathrm{chan}}, E_{\mathrm{pho}} \right) =\frac{\mathrm{Norm}}{\sigma \sqrt{2 \pi}} e^{-\frac{ (E_{\mathrm{chan}} - E_{\mathrm{pho}} )^{2} }{2 \sigma^{2}}} \xrightarrow{E_{\mathrm{chan}} = E_{\mathrm{pho}}} {RMF}_{\mathrm{max}} =\frac{\mathrm{Norm}}{\sigma \sqrt{2 \pi}} 
\end{aligned}
\label{eqn:RMF-shifting}
\end{equation}

    \item The shifting process widens the RMF distribution: This is actually to be expected and amounts to a factor of (1+redshift) similarly to energy shifting, as detailed by Equation \ref{eqn:RMF-widening}. 
    
\begin{equation}
\begin{aligned}
    \mathrm{Width}_{\mathrm{\ Original}} &= E_{\mathrm{\ high}} – E_{\mathrm{\ low}} \\
    \mathrm{Width}_{\mathrm{\ Shifted}} &= E_{\mathrm{\ high}} \cdot (1+z) – E_{\mathrm{\ low}} \cdot (1 + z) \\
    \mathrm{Width}_{\mathrm{\ Shifted}} &= (E_{\mathrm{\ high}} – E_{\mathrm{\ low}}) \cdot (1+z) \\
    \mathrm{Width}_{\mathrm{\ Shifted}} &= \mathrm{\ Width}_{\mathrm{\ Original}} \cdot (1+z) 
\end{aligned}
\label{eqn:RMF-widening}
\end{equation}

    \item As consequence of widening the RMF distribution the norm also increases by a factor of (1 + redshift) as shown by Equation \ref{eqn:RMF-norm}. 
    
\begin{equation}
\begin{aligned}
\mathrm{RMF}_{\mathrm{max}} =\frac{\mathrm{Norm}}{\sigma \sqrt{2 \pi}} &\xrightarrow{} \mathrm{Norm} = \mathrm{RMF}_{\mathrm{max}} \cdot \sigma \sqrt{2 \pi} \\
\sigma_{\mathrm{\ Shifted}}  = \sigma_{\mathrm{\ Original}} \cdot (1+z)  &\xrightarrow{} \mathrm{Norm}_{\mathrm{\ Shifted}}  = \mathrm{Norm}_{\mathrm{\ Original}} \cdot (1+z)
\end{aligned}
\label{eqn:RMF-norm}
\end{equation}

\item  Therefore it is necessary to re-normalize the distribution after shifting, resulting in a less peaked distribution, where the maximum value is reduced by a factor of 1/(1 + redshift) as shown by Equation \ref{eqn:RMF-renorm}. 

\begin{equation}
\begin{aligned}
\left( \mathrm{RMF}_{\mathrm{max}} \right)_{\mathrm{\ Shifted}} = \frac{\mathrm{Norm}}{\sigma_{\mathrm{\ Shifted}} \sqrt{2 \pi}} = \frac{\mathrm{Norm}}{\sigma \cdot (1+z) \sqrt{2 \pi}} = \frac{\left( \mathrm{RMF}_{\mathrm{max}} \right)_{\mathrm{\ Shifted}}}{1 + z}
\end{aligned}
\label{eqn:RMF-renorm}
\end{equation}

\end{itemize}

\begin{figure}[ht!]
    \centering
    \includegraphics[width=0.7\textwidth]{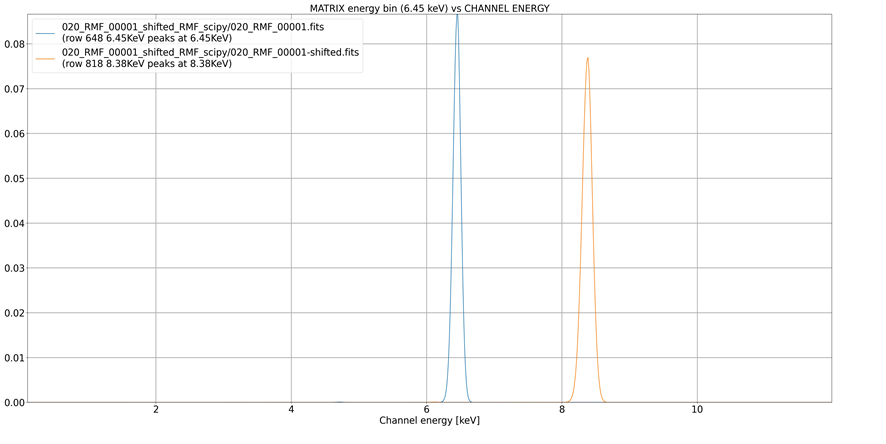}
    \caption { \footnotesize {Example of RMF shifting at z=0.3, for the case of incident photon at 6.7keV in the observe frame. Notice that before shifting (blue) the RMF peaks along channel bins at the same energy as the incident photon (6.7keV), and after shifting (orange) it peaks at the shifted energy $8.1 \mathrm{keV} = 6.7 \mathrm{keV} \cdot (1 + 0.3)$ } }
     \label{fig:RMF-shifting}
\end{figure}

\begin{figure}[ht!]
    \centering
    \includegraphics[width=0.7\textwidth]{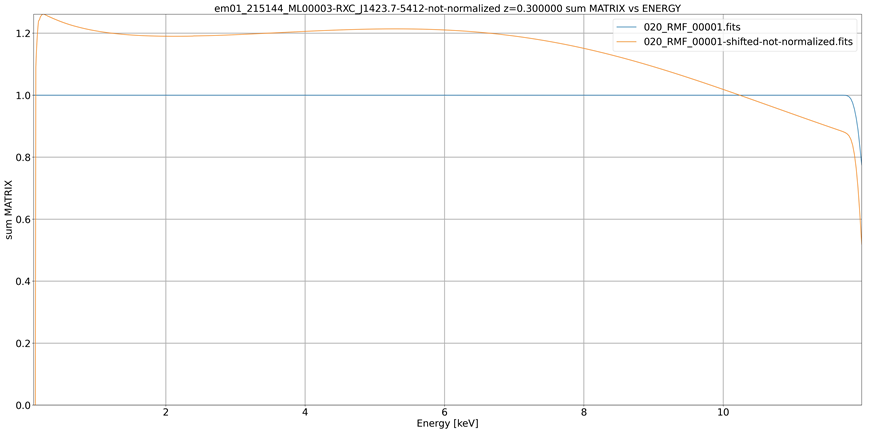}
    \caption { \footnotesize { Example of normalization after shifting at z=0.3. Notice that without normalization there would be a non-flat response across energies after shifting (orange), however the original norm before shifting is flat across energies (blue)} }
    \label{fig:RMF-norm}
\end{figure}

In summary the resulting RMF distribution, is shifted, widened and less peaked after normalization. Figure \ref{fig:RMF-shifting} shows the example of RXC J1423.7-5412, a high red-shift cluster at z=0.3, for the case of incident photon at 6.7keV in the observe frame. The re-normalization process guarantees a flat response across energies as shown in Figure \ref{fig:RMF-norm}

\section{Validation of the shifting procedure} \label{sec:Validation}

\begin{figure}[ht!]
    \centering
    \includegraphics[width=1.0\textwidth]{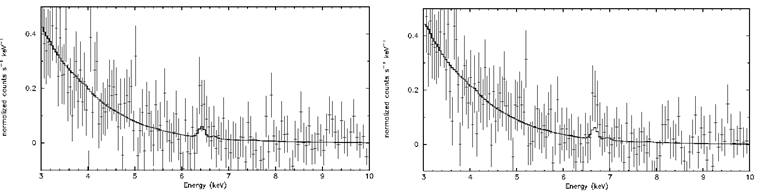}
    \caption { \footnotesize { Example of fitting a shifted spectra at z=0.051 (A3571). The 6.7-keV K$\alpha$ complex appears at 6.37keV in the observed frame (left), and after shifting it appear at 6.7keV in the source frame, with the fit model matching the location properly (right).} }
    \label{fig:A3571-shift}
\end{figure}

The first validation test of the shifting procedure simply consist of fitting the shifted spectra and make sure that the modelled line emission emerges at the expected energies. For example the The 6.7-keV K$\alpha$ complex is a strong line at high energy where the shift is more noticeable, therefore it is a good line to perform this test. Figure \ref{fig:A3571-shift} shows the resulting fit in the hard band, before and after shifting for a cluster at z=0.051 (A3571). The 6.7-keV K$\alpha$ complex appears at 6.37keV in the observed frame, and after shifting it appear at 6.7keV in the source frame, with the fit model matching the location properly.

The second, more quantitative test, consist of comparing the fitted temperatures and abundances of the shifted and un-shifted spectra, to make sure they are not significantly changed as a result of the shifting procedure and therefore it does not affect the physical results. 

\begin{figure}[ht!]
    \centering
    \includegraphics[width=1.1\textwidth]{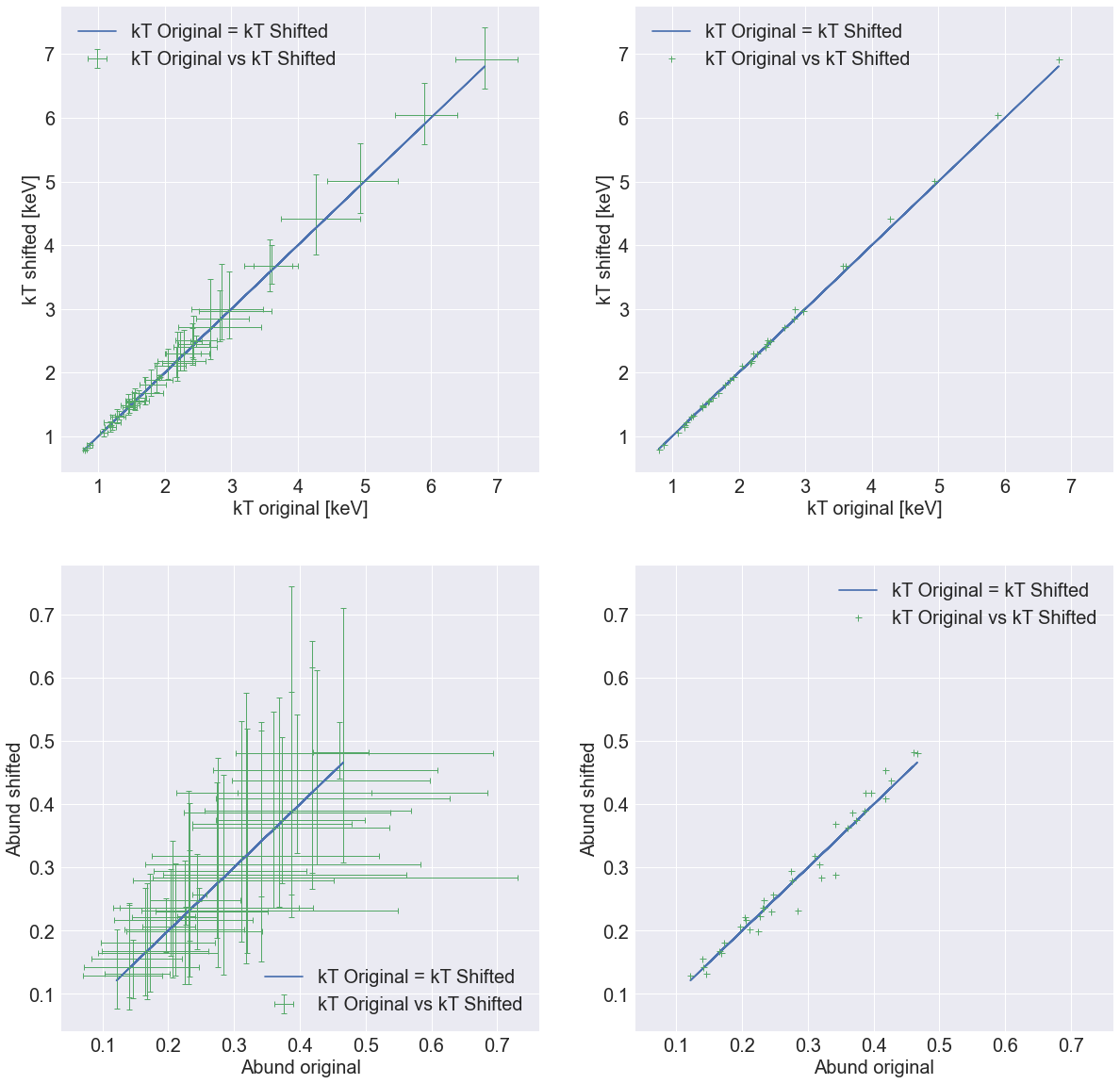}
    \caption { \footnotesize { Top: Comparison of the temperature measurements. Bottom: Comparison of the abundance measurements. For all panels the X axis corresponds to unsifted spectra measurements, and the Y axis corresponds to shifted spectra measurements. Left panels shows the comparison with error bars, and right panels without error bars to better notice the deviation. } }
    \label{fig:validation}
\end{figure}

As shown in Figure \ref{fig:validation} if the temperature / abundance measurements are well constrained in the original spectra (smaller error bars), then the same temperature / abundance measurements are re-obtained after shifting the spectra. 

However if the temperature / abundance measurements are not well in the original spectra (larger error bars), then the temperature / abundance measurements obtained after shifting the spectra deviate more from the original measurements. 

\section{Stacking} \label{sec:Stacking}

\begin{figure}[ht!]
    \centering
    \includegraphics[width=1.0\textwidth]{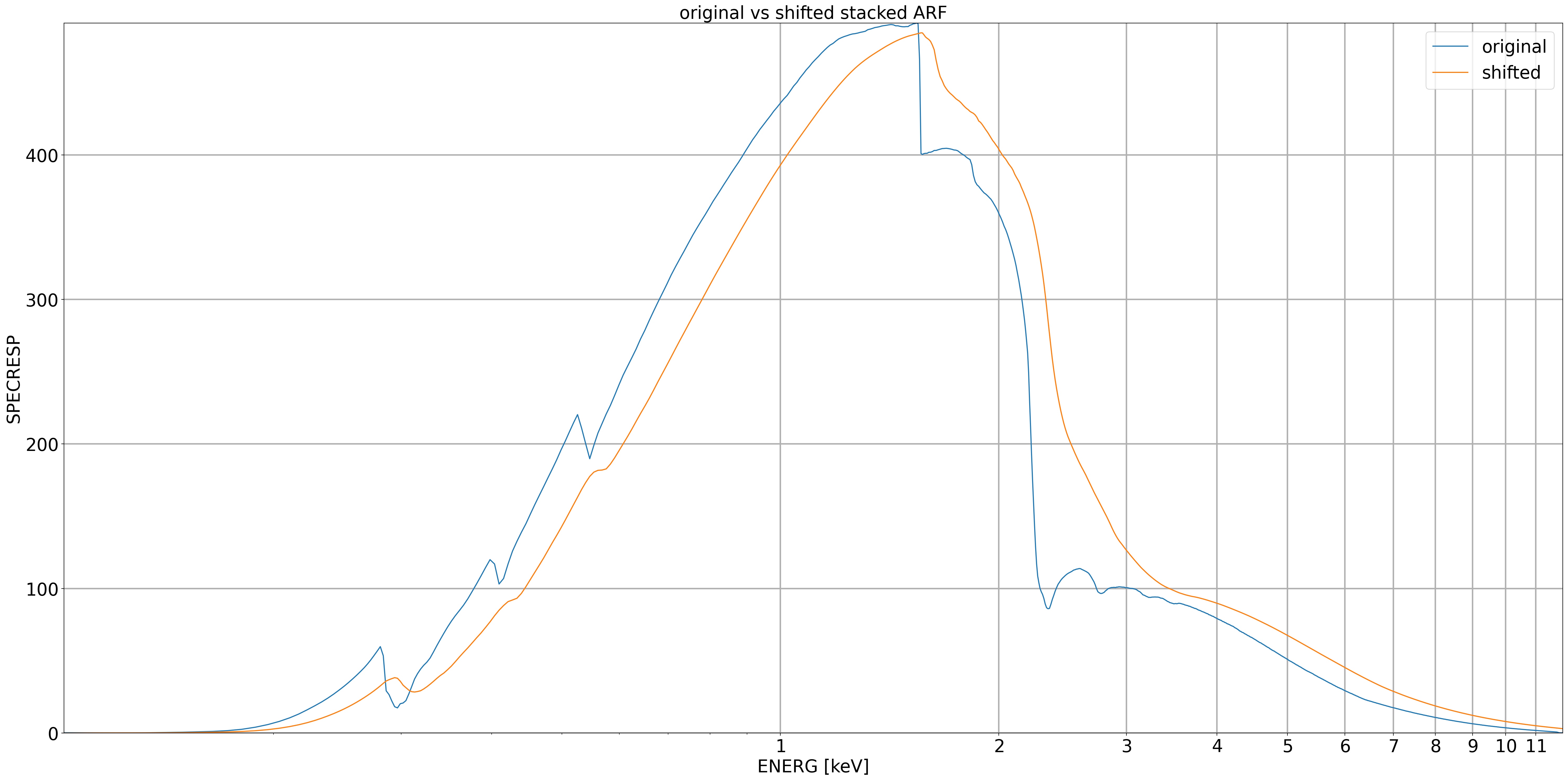}
    \caption { \footnotesize { Blue: Original effective area, Orange: Average effective area after stacking all clusters using as weights the total number of counts corrected by the background contribution. } }
    \label{fig:stacked_ARF}
\end{figure}

The stacking procedure consists of three parts:

\begin{itemize}

    \item In the first place source spectra are stacked without using any weights, that is by simply addition. Notice that the counts themselves weight the contribution of each cluster, as they are directly proportional to the X-Ray flux. The spectra addition  is done using the mathpha tool from the HEASOFT package \citep{drake2016heasarc}. We applying POISS-1 statistics valid for low counts, following the algorithm described in \cite{gehrels1986confidence}. The BACKSCALE keyword is set to 1.0 as prescribed by the ASCA reduction guide \citep{Ebisawa1997Background}.
    
    \item Secondly we stack background spectra also via mathpha and POISS-1 statistics, but in this case using a weighting schema to account for the exposure and background scaling as described in the ASCA reduction guide \citep{Ebisawa1997Background}: If $X_1 , X_2, .., X_n$ and $Y_1 , Y_2, .., Y_n$ are the EXPOSURE and BACKSCALE keywords of the source spectra $1, 2, .., n$, and $X_1^{'} , X_2^{'}, .., X_n^{'}$ and $Y_1^{'} , Y_2^{'}, .., Y_n^{'}$ the EXPOSURE and BACKSCALE keywords of the corresponding background spectra $1, 2, .., n$, then the weights to be used for the background stacking are $C_1 , C_2, .., C_n$ given by Equation \ref{eqn:weights}, and the BACKSCALE keyword of the stacked background spectra is given by Equation \ref{eqn:backscale}:

\begin{equation}
\begin{aligned}
C_{1}=\frac{X_{1}^{\prime}+X_{2}^{\prime}+,,,+X_{n}^{\prime}}{X_{1}\left(Y_{1} / Y_{1}^{\prime}\right)+X_{2}\left(Y_{2} / Y_{2}^{\prime}\right)+,,+X_{n}\left(Y_{n} / Y_{n}^{\prime}\right)}\left(\frac{X_{1}}{X_{1}^{\prime}}\right)\left(\frac{Y_{1}}{Y_{1}^{\prime}}\right) \\
C_{2}=\frac{X_{1}^{\prime}+X_{2}^{\prime}+,,,+X_{n}^{\prime}}{X_{1}\left(Y_{1} / Y_{1}^{\prime}\right)+X_{2}\left(Y_{2} / Y_{2}^{\prime}\right)+,,+X_{n}\left(Y_{n} / Y_{n}^{\prime}\right)}\left(\frac{X_{2}}{X_{2}^{\prime}}\right)\left(\frac{Y_{2}}{Y_{2}^{\prime}}\right) \\
C_{n}=\frac{X_{1}^{\prime}+X_{2}^{\prime}+,,,+X_{n}^{\prime}}{X_{1}\left(Y_{1} / Y_{1}^{\prime}\right)+X_{2}\left(Y_{2} / Y_{2}^{\prime}\right)+,,+X_{n}\left(Y_{n} / Y_{n}^{\prime}\right)}\left(\frac{X_{n}}{X_{n}^{\prime}}\right)\left(\frac{Y_{n}^{\prime}}{Y_{n}^{\prime}}\right) .
\end{aligned}
\label{eqn:weights}
\end{equation}

\begin{equation}
\begin{aligned}
(B A C K S C A L)_{b g d}=\frac{X_{1}+X_{2}+,,,+X_{n}}{X_{1}\left(Y_{1} / Y_{1}^{\prime}\right)+X_{2}\left(Y_{2} / Y_{2}^{\prime}\right)+,+X_{n}\left(Y_{n} / Y_{n}^{\prime}\right)} 
\end{aligned}
\label{eqn:backscale}
\end{equation}

    \item Finally the ARF and RMF are stacked using the addarf and addrmf tools from the HEASOFT package. In this case we use as weights the total number of counts corrected by the background contribution, so that the stacked ARF and RMF follows the contributions of each cluster in the stacked spectra. The resulting stacked RMF is quite smooth as shown in Figure \ref{fig:stacked_ARF}.
    
\end{itemize}

\section{Final stack selection and filters} \label{sec:Filters}

For our final stack we apply the following filters:

\begin{itemize}

    \item First of all we don't use the data from TM5 and TM7 to prevent the problems derived from the light leak as described in section \ref{sec:lightLeak}. 
    
    \item Also we exclude from our analysis the lowest count clusters, for which it is not even possible to constrain the temperature. This reduces the number of clusters from 1363 to 1255.
    
    \item We limit the highest red-shift to z=0.6. In the analysis of \cite{bulbul2014detection} this was limited to z=0.3, but for this work, we have a significant portion of counts in between z=0.3 and z=0.6, namely 11\% of the total data, corresponding to 458 clusters. Moreover including these clusters does not only increase the total amount of data, but also the benefits of smearing instrumental features thanks to a wider red-shift range.
    
    \item On the other hand our highest red-shift limit of z=0.6 also sets the lower limit of the spectral analysis, from the original $0.3 \mathrm{keV} $ limit of eROSITA to $0.3 \cdot (1 + 0.6) = 0.48 \mathrm{keV} $. The lower limit still allows to include Ca XVIII lines in the band [0.613-0.663] keV, and Ar XVI in the band [0.483-0.527] keV. The inclusion of these lower energy Ca/Ar lines allows to properly constrain the abundances of these Ca/Ar, which have prominent emission in the [3-4]keV band as well. In particular Ar XVIII lines at [3.323-3.318] keV and Ca XIX lines at [3.861-3.902] keV.
    
    \item Along the mass range we set the lower mass limit to $\mathrm{M}_{500} > 0.5 \mathrm{M}_{\odot} $.  In the analysis of \cite{bulbul2014detection} this was limited to $\mathrm{M}_{500} > 0.5 \mathrm{M}_{\odot} $,  but for this work, we have a significant portion of counts in between z=0.3 and z=0.6, namely 6.4\% of the total data, corresponding to 63 clusters. These clusters are in the boundary between galaxy groups and galaxy clusters, but it is interesting to include them in the analysis as they are colder clusters, that could potentially host cold gas, necessary ingredient for the charge exchange process.
    
\end{itemize}

After applying all these filters our final cluster stack comprises 1138 clusters, totalling 430649 counts. Including Virgo would increase the count number by ~120000 counts up to 550828 counts, however we exclude it to highlight what can be done by stacking only, when the individual spectra do not have many counts.



%% file: SpectralAnalysis.tex
\chapter{Spectral Analysis} \label{sec:SpectralAnalysis}

\section{Model selection and free parameters} \label{sec:ModelSelection}

Since we have a large sample of clusters, it is not possible to model the spectra with a single temperature APEC/APED component. \cite{bulbul2014detection} used four temperature components, but their stack was concentrated in 73 clusters, however in our case we have 1138 and a spread temperature distribution as shown in Figure \ref{fig:kT-z-hist}.

For modelling a wide temperature distribution the most suitable model in Xspec \citep{arnaud1996xspec} is the vgadem model, a multi-temperature plasma emission model, built on top of the APEC code, with Gaussian distribution of emission measure as shown in Equation \ref{eqn:vgadem}, where $Y_0$ is the total, integrated emission measure:

\begin{equation}
    Y(T)=\frac{Y_{0}}{\sqrt{2 \pi} \sigma_{T}} e^{-(T-T_0)^{2} / 2 \sigma_{T}^{2}}
    \label{eqn:vgadem}
\end{equation}

In reality it would be more appropriate to use a lognormal distribution, as the temperature distribution shown in Figure \ref{fig:kT-z-hist} is lognormal, and the hydrodynamic cosmological simulations show a lognormal distribution for both galaxies, and cluster of galaxies \citep{kawahara2007radial}, but unfortunately the lognormal emission model is not available in Xspec.

\begin{figure}[ht!]
    \centering
    \includegraphics[width=0.9\textwidth]{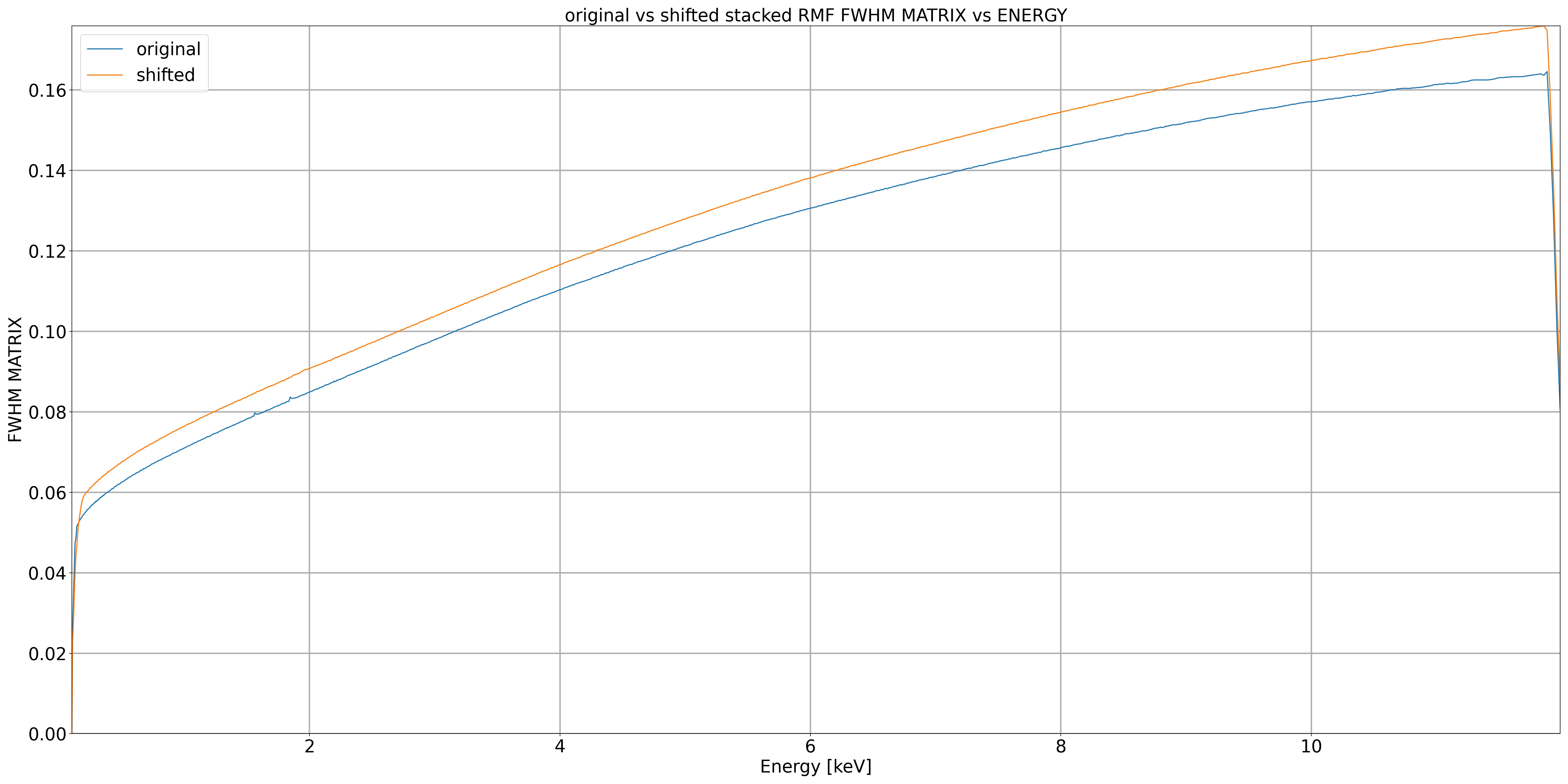}
    \caption { \footnotesize { Full Width at Half Maximum (FWHM) for eROSITA, obtained from the Matrix Redistribution File (RMF). Blue line indicates the original FWHM, and the orange line the FWHM resulting after shifting and stacking the sample of 1138 selected clusters. As explained in section \ref{sec:RMF} the shifting processes broadens the FWHM. } }
    \label{fig:RMF-FWHM-vs-ENERGY}
\end{figure}

On the other hand it is not necessary to consider velocity/turbulence broadening models (for example Xspec bapec models), because of the fact that the spectral resolution of eROSITA is not fine enough to measure the velocity dispersion. For example, turbulence broadening is usually around 100-500 km/s, which is equivalent to 0.01-0.02keV at 6.7keV according to Equation {ref:velocity-broadening}, clearly below the spectral resolution of eROSITA at that energy as shown by Figure \ref{fig:RMF-FWHM-vs-ENERGY}.

\begin{equation}
    \delta E = E \cdot z = E \cdot \left ( \sqrt{\frac{1+\frac{v_{\|}}{c}}{1-\frac{v_{\|}}{c}}} - 1 \right)
    \label{eqn:velocity-broadening}
\end{equation}

Finally, and as described in \ref{sec:APEC} it is necessary to take into account the absorption by the Inter Stellar Medium (ISM). For this Xspec offers the multiplicative model tbabs which implements the Tuebingen-Boulder ISM absorption model \citep{wilms2000absorption}, however we use the photoelectric absorption cross sections with variable abundances from \citep{balucinska1992photoelectric}. One problematic aspect of modeling the ISM absorption is that this is best done in the observed frame, before shifting, however most of the absorption happens in the soft band, where shifting has a smaller impact.

In the xspec terminology we adopt the tbabs*vgadem model, where the composed model consists of 11 free parameters as follows:

\begin{itemize}

    \item Galactic Hydrogen Column: This is the main parameter of the absorption model. Since we are modelling a stacked spectra we let hydrogen column free, to obtain an average effective hydrogen column.
    
    \item Temperature: This is the mean temperature for Gaussian emission measure distribution.
    
    \item Sigma Temperature: This is the sigma temperature for Gaussian emission measure distribution.
    
    \item Metal abundances: Metal abundances with respect to solar, and based on abundance tables from \cite{asplund2009chemical}. We let individual abundance of the following elements free: O, Ne, Mg, Si, Ar, Ca, Fe. For the case of Helium we fix it to the cosmic abundance, since otherwise it becomes degenerate with the norm of the Bremsstrahlung  component, provided that an increased helium abundance increases the Bremsstrahlung efficiency. The abundances of the remaining (trace) elements are linked to Fe.
    
    \item Norm: The model norm depends determines the total X-Ray flux, which depends on primarily on the average redshift and electron density of the clusters included in the stack.
    
\end{itemize}

\section{Temperature interpolation} \label{sec:TemperatureInterpolation}

\begin{figure}[ht!]
    \centering
    \includegraphics[width=0.6\textwidth]{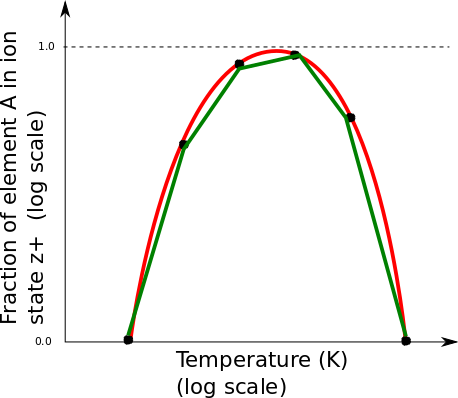}
    \caption { \footnotesize { Underestimation of ionization ratios due to interpolation: Since the ionization curve is concave, the interpolated value is systematically bias toward lower values.  Original from \cite{Foster2019Interpolation} } }
    \label{fig:Interpolation-IonFraction}
\end{figure}

\begin{figure}[ht!]
    \centering
    \includegraphics[width=0.6\textwidth]{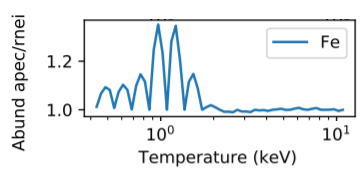}
    \caption { \footnotesize { Abundance measurement bias depending on the interpolation temperature. The temperatures where the bias is 0 correspond to the the nodes of the temperature interpolation grid. Original from \cite{Foster2019Interpolation} } }
    \label{fig:Interpolation-Abund}
\end{figure}

Another factor that affects the spectra model fitting is the temperature interpolation schema. \cite{Mernier2020codeuncertainties} pointed out in a recent paper, that the temperature interpolation schema used for obtaining the ionization ratios and emissivities systematically underestimates the true ionization ratios in points far from the interpolation grid nodes as shown in Figure \ref{fig:Interpolation-IonFraction}, resulting in abundance discrepancies up to 15\% as shown in Figure \ref{fig:Interpolation-Abund}.

The problem is partly caused by a sparse temperature interpolation grid in atomdb, which consist of 51 temperatures from $10^4K$ to $10^9K$. But regardless of the the grid resolution, since the ionization curve is concave, the interpolated value is systematically bias toward lower values.

\begin{figure}[ht!]
    \centering
    \includegraphics[width=0.6\textwidth]{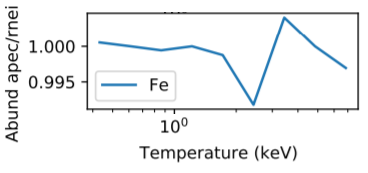}
    \caption { \footnotesize { Remaining abundance measurement bias, after introducing "on-the-fly" calculation of ionization ratios. Original from \cite{Foster2019Interpolation} } }
    \label{fig:Interpolation-Emissivity}
\end{figure}

Fortunately and as suggested by \cite{Foster2019Interpolation} this problem can be prevented, by using the same temperature interpolation schema for the abundance ratios as the non-equilibrium recombining collisional plasma code (rnei). In the rnei approach the ionization fraction is calculated on the fly, thus removing the interpolation bias over the ionization fractions. Interpolation of the emissivities  still suffers from temperature interpolation bias, but the remaining effect is much smaller, less than 1\% as shown in Figure \ref{fig:Interpolation-Emissivity}.

In practical terms, within Xspec it is only necessary to xset the APECUSENEI option to "yes" so that the calculation of ion fractions is routed through the NEI code. Additionally the newer version of atomb db (v 3.0.9) provides a 201 temperature interpolation grid.

\section{Flagging the gold edge} \label{sec:Flagging}

\begin{figure}[ht!]
    \centering
    \includegraphics[width=1.0\textwidth]{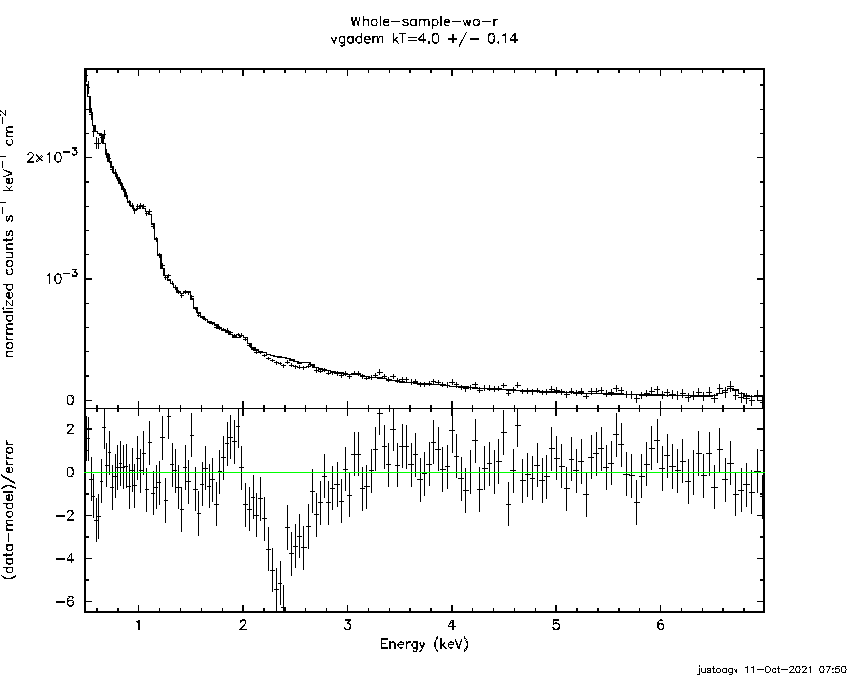}
    \caption { \footnotesize { Spectra corresponding to the entire stack, comprising 1138 clusters, modelled with a Gaussian distribution of temperatures (Tbasbs*vgadem). Notice the important residuals up to $4 \sigma $ around the gold edge, in the [2.1-2.7]keV range. We have a applied a 4-binning factor for visualization purposes. } }
    \label{fig:spectra-original}
\end{figure}

We first perform a quick fit test of our Tbasbs*vgadem model, for the case of the stacked spectra comprising 1138 clusters as described in Section \ref{sec:Filters}. For this test we use the Levenberg-Marquardt algorithm based on the CURFIT routine from Bevington. The test reveals important residuals up to $4 \sigma $ around the gold edge, in the [2.1-2.7]keV range as shown in Figure \ref{fig:spectra-original}..

As explained in section \ref{sec:calibration}, there are also important residuals above the 10\% level seen in the eROSITA calibration target A3266 \citep{sanders2021studying}. However these residuals are present even when extracting the core 6" arcmin spectra of A3266, and are not fixed by including the gold edge in the Vignetting function, as explained in Section \ref{sec:vignetting}.

\begin{figure}[ht!]
    \centering
    \includegraphics[width=1.0\textwidth]{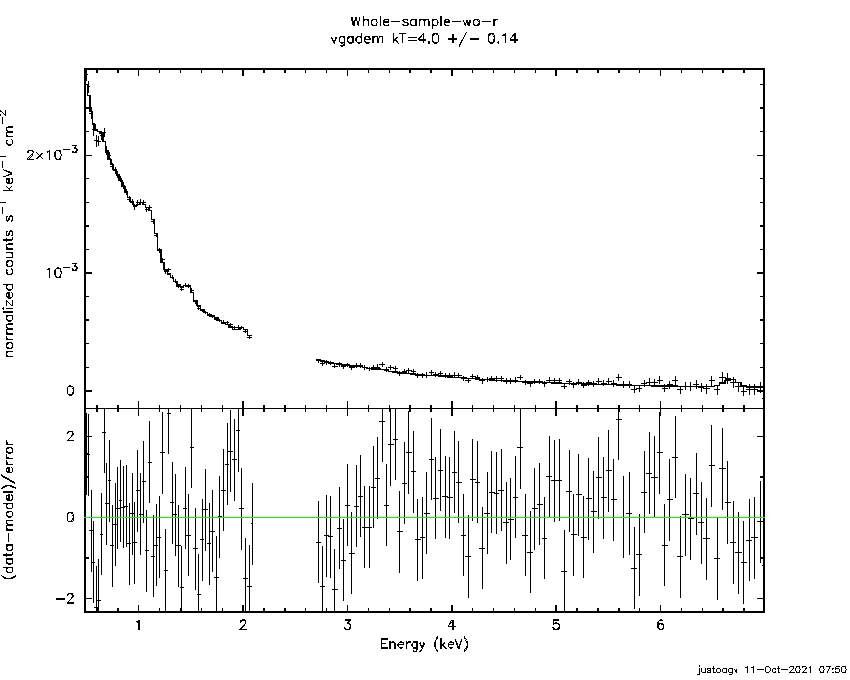}
    \caption { \footnotesize {  Spectra corresponding to the entire stack, comprising 1138 clusters, modelled with a Gaussian distribution of temperatures (Tbasbs*vgadem). In this case we have flagged the range corresponding to the shifted and smeared gold edge, in the [2.1-2.7]keV range. We have a applied a 4-binning factor for visualization purposes. } }
    \label{fig:spectra-flagged}
\end{figure}

Still the residuals are more prominent near the edges of the effective area function, with the gold edge standing as the most prominent, in the [2.0-2.4]keV range. However, since the shifting and stacking process smears the instrumental features, the residuals are seen in a relatively shifted and wider [2.1-2.7]keV range. We have no other option but to flag this range for the spectral analysis as shown in Figure \ref{fig:spectra-flagged}.

\section{Markov Chain configuration and priors}  \label{sec:MCMC}

Since our tbas*vgadem model consist of 11 free parameters (nH, temperature, sigma temperature, and O, Ne, Mg, Si, Ar, Ca, Fe abundances), we need a robust fit method to avoid local minimum. 

In this sense Xspec offers two types of Markov Chains Monte Carlo (MCMC), one based on the Metropolis-Hastings algorithm, which requires a choice of proposal distribution, however finding the best distribution can be difficult and makes MCMC harder to use.

Alternatively, the Goodman-Weare algorithm does not require a choice of proposal distribution. It works by running multiple sets of parameters, called walkers, which are re-generated for each step of the chain using the walkers from the current step. Therefore for this work we use Goodman-Weare MCMC.

However, to speed up convergence we do a pre-fit using the Levenberg-Marquardt algorithm based on the CURFIT routine from Bevington, and use an initial Gaussian prior with co-variances based on it (chain proposal gaussian fit)

\begin{figure}[ht!]
    \centering
    \includegraphics[width=1.0\textwidth]{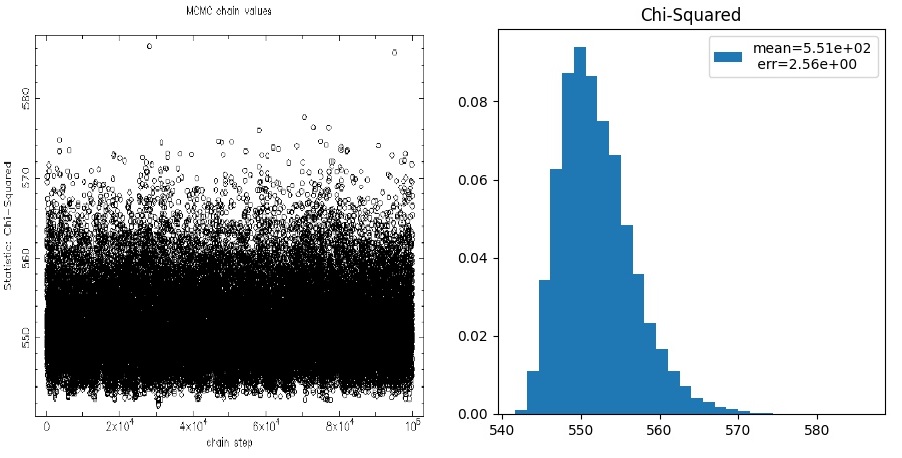}
    \caption { \footnotesize { $\chi^2$ statistic for each step of the chain, and the corresponding distribution, which shows no local minimum.} }
    \label{fig:mcmc-Chi-Squared}
\end{figure}

To be sure of convergence, we have experimented with the length of the burn phase, to make sure that the chain has reached a steady state. In our tests this is reached with a 200,000 step burn phase. The chain length is set to 100,000 steps, and we use 10 walkers, so the effective chain length is 1,000,000 steps. Figure \ref{fig:mcmc-Chi-Squared} shows the $\chi^2$ statistic for each step of the chain, and the corresponding distribution, which shows no local minimum.

\section{Fit result and constrains} \label{sec:Fit}

\begin{figure}[ht!]
    \centering
    \includegraphics[width=1.0\textwidth]{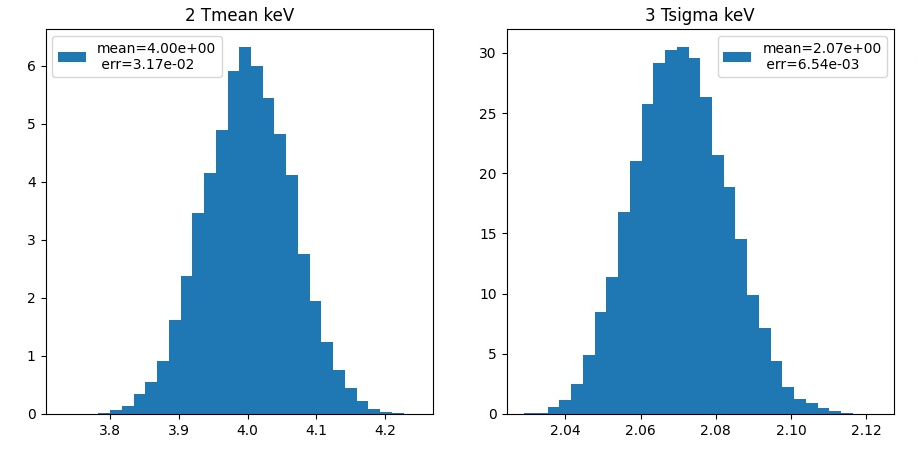}
    \caption { \footnotesize { MCMC histograms for the mean and sigma values of the Gaussian temperature distribution.} }
    \label{fig:Tmean-Tsigma}
\end{figure}

Via the MCMC Goodman-Weare algorithm presented in Section \ref{sec:MCMC} we can also obtain constrains for the mean and sigma values of the Gaussian temperature distribution obtaining $T_{\mathrm{mean}} = ( 4.00 \pm 0.03) K$ and $T_{\mathrm{sigma}} = (2.070 \pm 0.006) K$ as shown in Figure \ref{fig:Tmean-Tsigma}.

\begin{figure}[ht!]
    \centering
    \includegraphics[width=0.9\textwidth]{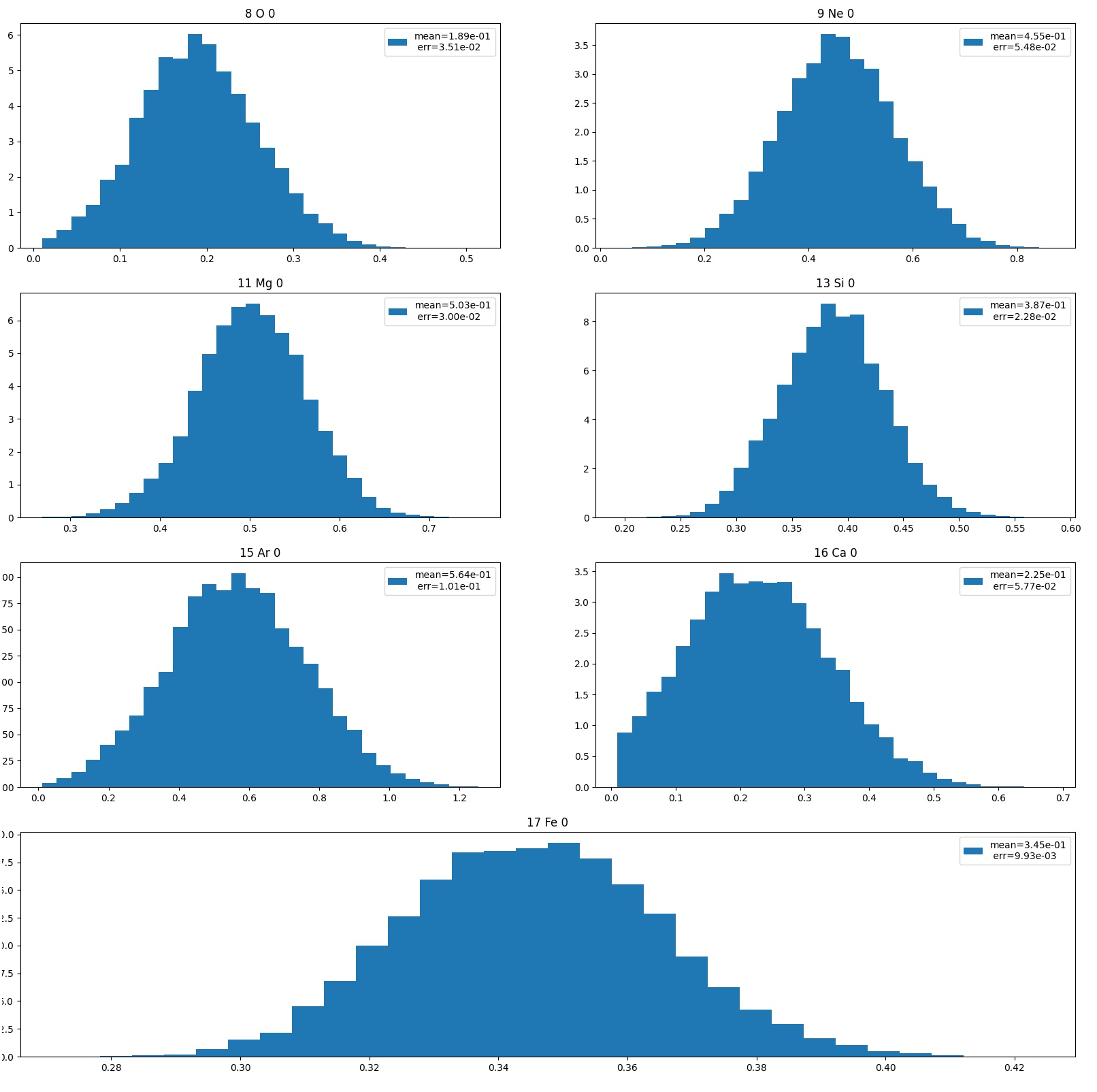}
    \caption { \footnotesize { MCMC histograms the individual abundances of O, Ne, Mg, Si, Ar, Ca and Fe. Notice the higher abundances of Ar and Ca in comparison with other elements, also linked with higher error. } }
    \label{fig:abundance}
\end{figure}

Additionally we can obtain constrains for the individual abundances of O $( 0.189 \pm 0.035)$, Ne $( 0.455 \pm 0.055)$, Mg $( 0.503 \pm 0.030)$, Si $( 0.387 \pm 0.023)$, Ar $( 0.564 \pm 0.10)$, Ca $( 0.225 \pm 0.58)$ and Fe $( 0.345 \pm 0.0.001)$ as shown in Figure \ref{fig:abundance}. We do notice the higher abundances of Ar and Ca in comparison with other elements, also linked with higher error. This can be an indication of X-Ray excess in the [3-4]keV band, where the Ar XVIII lines at [3.323-3.318] keV and Ca XIX lines at [3.861-3.902] keV are located.

\begin{figure}[ht!]
    \centering
    \includegraphics[width=1.0\textwidth]{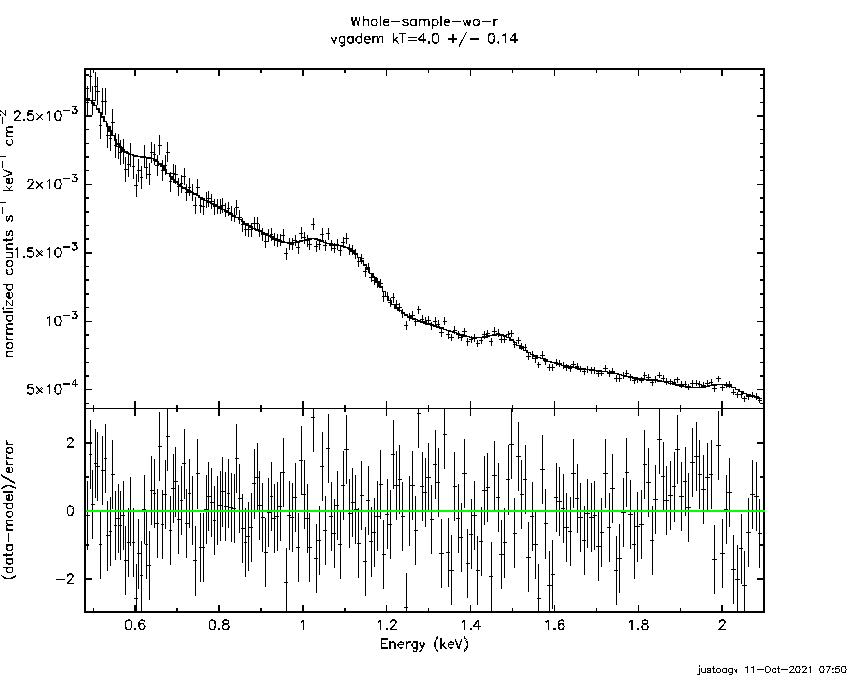}
    \caption { \footnotesize { Soft band ([0.48-2.1]keV) spectra corresponding to the entire stack, comprising 1138 clusters, modelled with a Gaussian distribution of temperatures (Tbasbs*vgadem). Notice that the residuals barely exceed $\pm 2 \sigma$. } }
    \label{fig:spectra-soft-bin-1}
\end{figure}

\begin{figure}[ht!]
    \centering
    \includegraphics[width=1.0\textwidth]{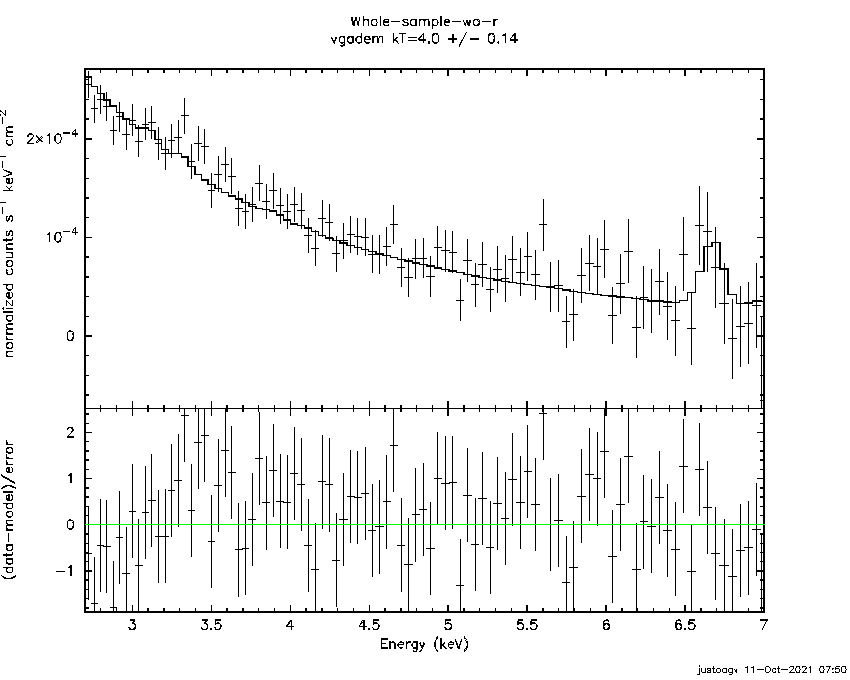}
    \caption { \footnotesize { Hard band ([2.7-7.0]keV) spectra corresponding to the entire stack, comprising 1138 clusters, modelled with a Gaussian distribution of temperatures (Tbasbs*vgadem). Notice that the residuals barely exceed $\pm 2 \sigma$, although there are noticeable outliers in the [3-4]keV band, which seem to be centered around 3.5keV and decrease at lower and higher energies. We have a applied a 4-binning factor for visualization purposes.  } }
    \label{fig:spectra-hard-bin-4}
\end{figure}

Finally we present the fit results in Figures \ref{fig:spectra-soft-bin-1} (soft band [0.48-2.1]keV) and \ref{fig:spectra-hard-bin-4} (hard band [2.7-7.0]keV). In both cases the residuals barely exceed $\pm 2 \sigma$, although there are noticeable outliers in the [3-4]keV band, which seem to be centered around 3.5keV and decrease at lower and higher energies. 

\clearpage
 
\section{Charge Exchange Model}  \label{sec:CX}

Now we proceed to add the charge exchange (CX) component to our model, to see if the residuals (outliers) near 3.5keV could be explained with an additional CX spectral component. For this we employ the ACX model by \cite{smith2012approximating}, which is nicely integrated in Xspec, and we can include as an additive component resulting in a total Tbabs*(vgadem + vacx) model

As described in \cite{smith2012approximating} ACX is actually an approximation to the X-ray spectrum emitted from astrophysical charge exchange. A complete charge exchange model requires a vast number of atomic calculations, but ACX relies on an approximate calculation of the cross sections. Namely the method described by \cite{wegmann1998x} which uses a hydrogenic model for the CX cross section into the highly excited state, as shown in Equation \ref{eqn:cx-cross-section} where q is the charge of the ion, n the principal quantum number, and $I_p$ is the ionization potential in atomic units (27.2 eV)

\begin{equation}
    \sigma=8.8 \times 10^{-17} \frac{q-1}{\frac{q^{2}}{2 n^{2}}-\left|I_{\mathrm{p}}\right|} \quad\left(\mathrm{cm}^{2}\right)
    \label{eqn:cx-cross-section}
\end{equation}

To predict the actual line emission it is required to first determine the atomic level distribution of the charge-exchange ion. For this ACX uses the approximation described in \cite{janev1985state} which obtained that the peak of the principle quantum number n distribution is given by Equation \ref{eqn:cx-n-peak}, where q is the charge of the ion, $I_H$ is the ionization energy of the neutral ion (assumed here to be hydrogen), and $I_P$ is the ionization potential in atomic units (27.2 eV):

\begin{equation}
    n^{\prime}=q \sqrt{\frac{I_{\mathrm{H}}}{I_{\mathrm{p}}}}\left(1+\frac{q-1}{\sqrt{2 q}}\right)^{-1 / 2}
    \label{eqn:cx-n-peak}
\end{equation}

The default model of ACX (model 8) calculates $n^{\prime}$ and applies a weighted distribution, so that if for example $n^{\prime} = 4.7$ then 30\% of the ions would populate n = 4 while 70\% would be in n = 5.

The cross section is summed over all LS states for a given n, but the primary uncertainty in the process is the total final angular momentum (L) of the exchanged electron, which is velocity-dependent, and this difficult to constrain. ACX offers different weighting schema, but the default model (number 8) evenly weights the total angular momentum L.

In summary, for our analysis we use the default ACX model (number 8) which applies a weighted distribution for the peak of the principle quantum number n, and an even distribution for the the total angular momentum L, without adding a dependency on the velocity. It is clear that this is just an approximation, but we simply aim to estimate the potential contribution of CX in our spectrum.

ACX is actually offered with a VACX variant, which allows to individually set the abundances of each element, therefore for our analysis we link the abundances of all elements in the ACX model to those of the VAPEC model. The temperature of the ACX model refers to the ion temperature (not that of the cold gas) and is linked to the average temperature of the Gaussian distribution of emission measure. Therefore the only free parameter added by including the ACX component is the normalization factor. 

\begin{figure}[ht!]
    \centering
    \includegraphics[width=1.0\textwidth]{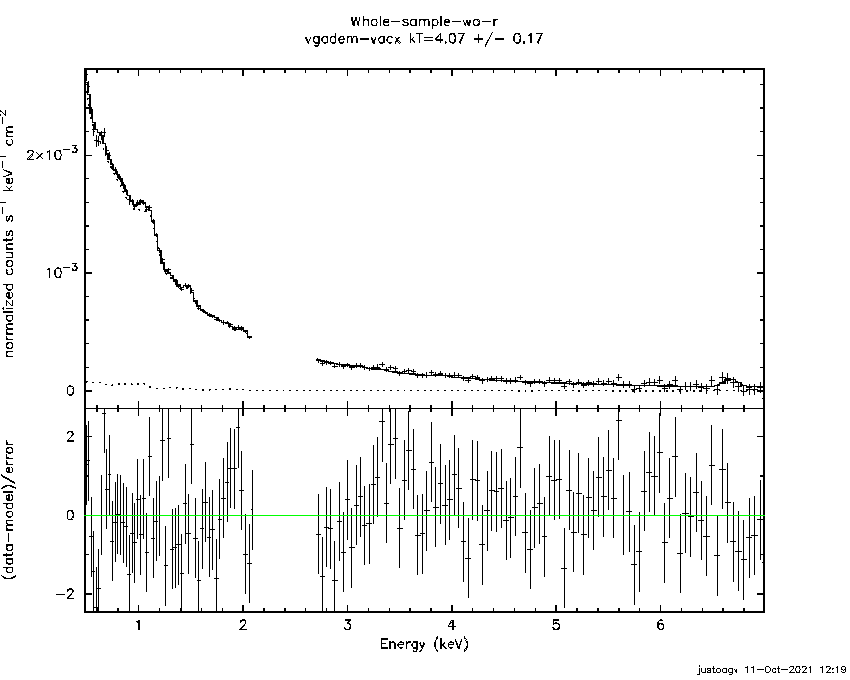}
    \caption { \footnotesize {  Spectra corresponding to the entire stack, comprising 1138 clusters, modelled with Tbasbs*(vgadem + vacx). The dashed line indicates the estimated charge exchange contribution to the spectra according to the ACX model. We have a applied a 4-binning factor for visualization purposes. } }
    \label{fig:spectra-cx}
\end{figure}

Then we repeat the MCMC fit procedure described in section \ref{sec:Fit} with our Tbabs*(vgadem + vacx) model. The results show an almost negligible contribution of CX to the stacked spectra, as indicated by the dashed line shown in Figure \ref{fig:spectra-cx}. The contribution of charge exchange is at the 3\% level in the best case, for the lowest energies $< 1.1 \mathrm{keV}$.

\begin{figure}[ht!]
    \centering
    \includegraphics[width=1.0 \textwidth]{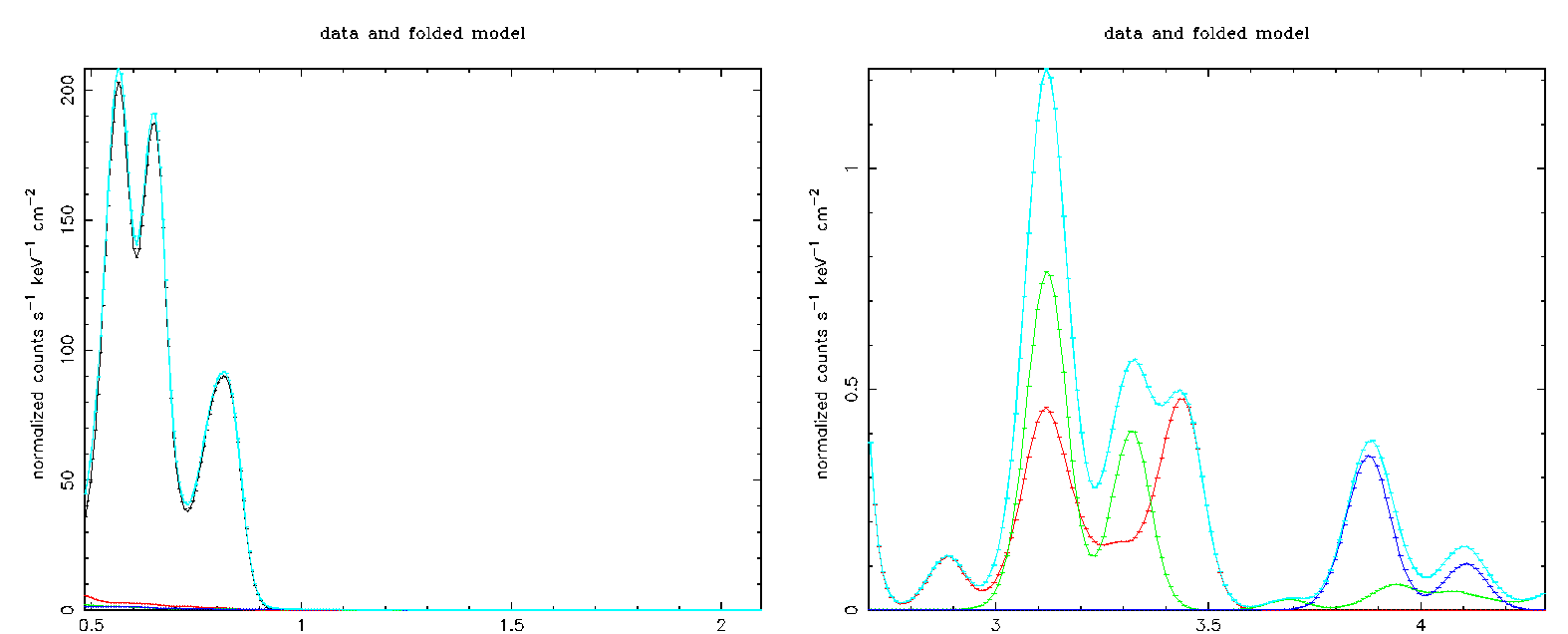}
    \caption { \footnotesize { Simulation of charge exchange spectra for eROSITA based on ACX for 3.5keV, and abundance at 0.3. The left panel shows the soft band [0.48-2.1]keV with the prominent OVII 0.56keV CX line, with a flux of $200\ \mathrm{counts} / \mathrm{s} \cdot \mathrm{cm}^2 \cdot \mathrm{keV}$ and the right panel the hard band [0.48-2.1]keV, with SXVI CX emission around 3.4–3.55 with a flux of $0.5\ \mathrm{counts} / \mathrm{s} \cdot \mathrm{cm}^2 \cdot \mathrm{keV}$, 400 orders of magnitude less. Cyan indicates the overall charge exchange emission, green Ar, red S, and blue Ca. } }
    \label{fig:spectra-cx-sim}
\end{figure}

A  more detailed analysis of this result reveals that despite the excess around 3.5keV, which could correspond to charge exchange emission by bare sulfur ions, the lack of charge exchange emission by all other ions imposes hard constrains, on the overall charge exchange process. In particular the most prominent charge exchange lines, corresponding to OVII (0.56keV) is not present in the spectra. Notice that the OVII 0.56keV CX line is 400 orders of magnitude higher than the SXVI CX emission in the 3.4–3.55 band as shown in Figure \ref{fig:spectra-cx-sim}, a simulation of charge exchange spectra for eROSITA based on ACX for 3.5keV, and abundance at 0.3.

%% file: Conclusions.tex
\addcontentsline{toc}{chapter}{\protect Conclusions}
\chapter*{Conclusions} \label{Conclusions}

In summary with this work we have shown the capabilities of the shifting and stacking algorithm outlined by \cite{bulbul2014detection}. We are able to stack up to 1138 clusters, totalling 430649 counts, and obtain constrains for the individual abundances of O, Ne, Mg, Si, Ar, Ca and Fe. 

Excluding the residuals associated with the gold edge, which were also reported for the A3266 calibration target by \cite{sanders2021studying}, our residuals barely exceed exceed $\pm 2 \sigma$ across the spectra. However the amount of data available in eRASS-1 (1138 clusters, totalling 430649 counts) is still a factor of 10 less in comparison with the stacked spectra of \cite{bulbul2014detection}, since they obtained $3.2 \cdot 10^6$ counts for the MOS stack, and $2.1 \cdot 10^6$ counts for the PN stack. 

In the [3-4]keV range the residuals also barely exceed $\pm 2 \sigma$, however we do notice the higher abundances of Ar and Ca in comparison with other elements, also linked with higher error. This can be an indication of X-Ray excess in the [3-4]keV band, where the Ar XVIII lines at [3.323-3.318] keV and Ca XIX lines at [3.861-3.902] keV are located. Additionally the residuals in the [3-4]keV band are mostly positive, and seem to be centered around 3.5keV and decrease at lower and higher energies. 

One major problem for precision works of this kind, is that it is necessary to have a very good calibration. However, at the time being and according to \cite{sanders2021studying} the calibration residuals are above the 10\% level. In this work we have studied the impact of adding the gold edge in the vignetting function, but this does not solve the problems seen in A3266. Further improvements in the calibration are necessary, to reach a level that allows to reliably assess the nature of fit residuals.

On the other hand, we are able to rule out the charge exchange emission scenario proposed by \cite{gu2015novel} to explain the excess around 3.5keV. By excluding the data from TM5 and TM7 from our analysis we are able to use the lower end of eROSITA spectra range, down to 0.48keV. Therefore we can assess the presence of the prominent charge exchange line of OVII at 0.56keV, which unfortunately is not detected, and moreover the overall charge exchange emission across the spectra is only at the 3\% level.

The residuals in the [3.4–3.55]keV band are compatible with charge exchange emission from SXVI (bare sulfur ions) as stated by \cite{gu2015novel} , but charge exchange emission from OVII at 0.56keV should also be present, since it is 200 orders of magnitude higher than charge emission from SXVI in the [3.4–3.55]keV band. Nevertheless, the ACX model \citep{smith2012approximating} is only an approximation, and it can be that further improvements of the model increase the estimation of charge exchange emission in the spectra of galaxy clusters.

Future developments of this work should aim to first fix the calibration problems, secondly add more data, possibly by stacking eRASS-2,3 and 4, and finally improving the fit model. 

The fit model needs to be improved by correcting the ISM absorption before shifting, using a more appropriate emission measure distribution, namely a log-normal distribution, and using a more sophisticated charge exchange model, with measured cross-sections and velocity dependence. 

In this sense there are alternative X-Ray analysis packages such Spex \citep{kaastra2017spex}, which include log-normal emission measure distributions, and velocity dependence in the charge exchange process. This could reveal a higher contribution of charge exchange emission depending on the collision velocity as shown by \cite{gu2015novel}.

%% file: Bibliography.tex
\bibliographystyle{humannat}

\bibliography{Bibtex}

%% file: Declaration.tex
\chapter*{}

Declaration: 
\vspace{1cm}

I hereby declare that this thesis is my own work, and that I have not used any sources and aids other than those stated in the thesis. 

\vspace{2cm}

München, 03.01.2022
\vspace{2cm}  

\begin{figure}[ht!]
    \includegraphics[width=0.4\textwidth]{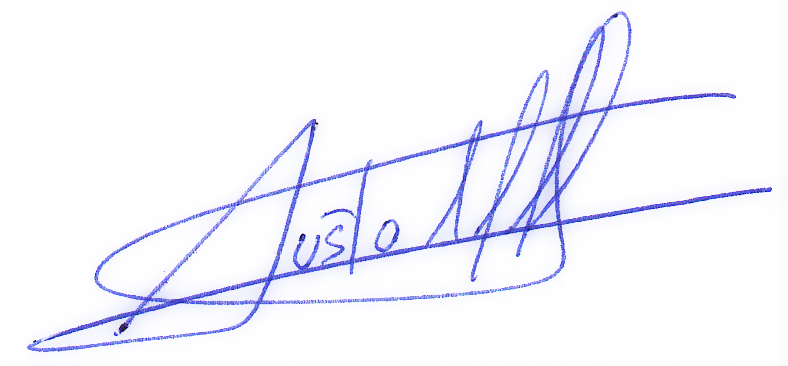}
\end{figure}
\rule{6cm}{0.4pt}

Justo Antonio Gonzalez Villalba